\documentclass[twocolumn]{svjour3}
\usepackage{amsmath}
\usepackage{amssymb}
\usepackage{url}
\usepackage[english]{babel}
\usepackage{graphicx}
\usepackage{siunitx}
\usepackage[utf8]{inputenc}
\usepackage[sort&compress,numbers]{natbib}
\usepackage{subfigure}
\usepackage{hyperref}

\begin{document}

\title{Massively parallel simulations for disordered systems}

\author{Ravinder Kumar \and
  Jonathan Gross \and
  Wolfhard Janke \and
  Martin Weigel
}

\institute{
  Ravinder Kumar \and Jonathan Gross \and Wolfhard Janke \at Institut f\"ur
  Theoretische Physik, Universit\"at Leipzig,
  IPF 231101, 04081 Leipzig, Germany
  \and
  Ravinder Kumar \and Martin Weigel \at
  Centre for Fluid and Complex Systems, Coventry University, Coventry CV1 5FB,
  United Kingdom \\\email{Martin.Weigel@complexity-coventry.org}
}

\maketitle

\begin{abstract}
  Simulations of systems with quenched disorder are extremely demanding, suffering
  from the combined effect of slow relaxation and the need of performing the disorder
  average. As a consequence, new algorithms, improved implementations, and
  alternative and even purpose-built hardware are often instrumental for conducting
  meaningful studies of such systems. The ensuing demands regarding hardware
  availability and code complexity are substantial and sometimes prohibitive. We
  demonstrate how with a moderate coding effort leaving the overall structure of the
  simulation code unaltered as compared to a CPU implementation, very significant
  speed-ups can be achieved from a parallel code on GPU by mainly exploiting the
  trivial parallelism of the disorder samples and the near-trivial parallelism of the
  parallel tempering replicas. A combination of this massively parallel
  implementation with a careful choice of the temperature protocol for parallel
  tempering as well as efficient cluster updates allows us to equilibrate
  comparatively large systems with moderate computational resources.
\end{abstract}

\section{Introduction}
\label{sec:intro}

Four or five decades of a concerted research effort notwithstanding, systems with
strong disorder such as spin glasses and random-field systems
\cite{binder:86a,young:book} are still puzzling researchers with a fascinating range
of rich behaviors that are only partially understood. Examples include the nature of
the spin-glass phase in low dimensions \cite{banos:12,wang:14}, universality and
dimensional reduction at critical points \cite{parisi:79a,fytas:13,fytas:16}, as well
as dynamic phenomena such as rejuvenation and aging
\cite{baity-jesi:18,baity-jesi:19}. While mean-field theory and perturbation
expansions for finite dimensions have set the stage for the field
\cite{nattermann:97,mezard:book}, a lot of the progress in recent years has been
through extensive numerical simulations, mostly in the form of Monte Carlo
simulations \cite{binder:book2} and ground-state calculations relying on
combinatorial optimization techniques \cite{hartmann:book}. While hence computational
methods have had a pivotal role in improving our understanding of systems with strong
disorder, simulations of such systems are far from technically straightforward. Due
to the rugged free-energy landscape \cite{janke:07} with a multitude of minima
separated by energy barriers standard approaches utilizing local updates such as
single-spin flip Metropolis or heat-bath algorithms are only able to equilibrate the
tiniest of samples, and generalized-ensemble techniques have more recently always
been used for simulating spin glasses, in particular, see, e.g.,
Refs.~\cite{berg:98a,houdayer:01,katzgraber:06a,banos:10,banos:12,sharma:11a,wang:15b}.

Parallel tempering or replica-exchange Monte Carlo has established itself as {\em de
  facto\/} standard for equilibrium simulations of spin-glass systems
\cite{hukushima:96a,wang:15}. The main difficulty there relates to the choice of the
temperature and sweep protocols for the method in order to achieve optimal (or at
least acceptable) mixing behavior of the overall Markov chain. A number of schemes to
this end have been proposed in the past
\cite{hukushima:99,katzgraber:06,bittner:08,rozada:19}. The basic methods use fixed
temperature sequences following (inversely) linear or exponential progressions, but
in many cases these lead to far from optimal performance. Adaptive approaches
\cite{hukushima:99,katzgraber:06,bittner:08} move the temperature points and set the
sweep numbers in such a way as to dynamically optimize the mixing behavior of the
chain, but these typically require rather extensive pre-runs to establish the best
parameters. Below we introduce and discuss a compromise approach that uses a family
of temperature protocols that can be optimized for the problem studied with moderate
effort. It is worthwhile noting that more recently an alternative to parallel
tempering known as population annealing \cite{hukushima:03,machta:10a} has gained
traction for simulations of disordered systems
\cite{wang:14,wang:15b,wang:15a,barash:18}, especially since it is particularly well
suited for parallel and GPU computing \cite{barash:16}. In the present paper,
however, we focus on the more traditional setup using replica-exchange Monte Carlo.

While parallel tempering has been able to speed up spin-glass simulations
dramatically, they still suffer from slow dynamics close to criticality and
throughout the ordered phase. Further relief could potentially be expected from
approaches alike to the cluster updates that have proven so successful for the
simulation of pure or weakly disordered systems
\cite{swendsen-wang:87a,wolff:89a,edwards:88a,chayes:98a}. While these methods can be
generalized quite easily to the spin-glass case \cite{swendsen:86,liang:92}, the
resulting algorithms typically result in clusters that percolate in the
high-temperature phase, way above the spin-glass transition, and hence such updates
are not efficient \cite{machta:07}. For the case of two dimensions, a cluster method
that exchanges clusters between pairs of replicas and thus operates at overall
constant energy turns out to be efficient if combined with local spin flips and
parallel tempering \cite{houdayer:01}. In three dimensions, however, also this
approach is affected by the early percolation problem, although recently a manual
reduction of the effective cluster size has been proposed as an {\em ad hoc\/} way of
alleviating this problem \cite{zhu:15a}. Similar cluster updates have also been
discussed for the case of systems without frustrating interactions, but in the
presence of random fields \cite{redner:98}.

Even with the best algorithms to hand relaxation times remain daunting, and with the
simultaneous presence of strong finite-size corrections to scaling the appetite of
researchers studying systems with strong disorder for more computing power seems
insatiable. As a result, enormous effort has also been invested in the optimization
of implementation details and the utilization of new hardware platforms. One line of
research, which is in the tradition of earlier hardware for spin systems
\cite{bloete:99a}, relates to the design and construction of special-purpose machines
based on field-programmable gate arrays (FPGAs) for simulations of spin glasses and
related problems \cite{belleti:09,baity-jesi:14a}. While this approach has been very
successful \cite{banos:12,baity-jesi:18,baity-jesi:19}, the financial and time effort
invested is enormous, and hence the demand for simpler, off-the-shelf solutions
remains strong. A significant competitor in this context are graphics processing
units (GPUs) that are able to deliver performances quite comparable to those of the
special-purpose machines to a much wider audience of users \cite{weigel:10a}. While
GPUs are more widely available and easier to program than FPGAs, many of the
approaches and code layouts proposed for efficient simulation of spin glasses on GPUs
are very elaborate, using a multitude of advanced techniques to speed up the
calculation \cite{bernaschi:10,yavorskii:12,baity-jesi:14,lulli:15}. In the present
paper, instead, we demonstrate how with very moderate effort and a straightforward
parallelization of pre-existing CPU code, excellent performance of spin-glass
simulations can be achieved on GPU.

The remainder of this paper is organized as follows. In Sec.~\ref{sec:methods} we
introduce the Edwards-Anderson spin glass and the parallel tempering method used for
its simulation. Subsequently, we discuss a new parameter-driven scheme for
determining an optimized temperature schedule. Finally, we shortly introduce a
cluster simulation method originally proposed for simulations of spin glasses in two
dimensions. Section \ref{sec:gpu} discusses the considerations relating to our GPU
implementation of this simulation scheme and how it relates to previous spin-glass
simulation codes on GPU. In Sec.~\ref{sec:performance} we benchmark the resulting
codes for the Ising spin glass in two and three dimensions, using discrete and
continuous coupling distributions. Finally, Sec.~\ref{sec:conclusion} contains our
conclusions.

\section{Model and methods}
\label{sec:methods}

\subsection{Edwards-Anderson spin glass}

While methods very similar to those discussed here can be used for simulations of a
wide range of lattice spin systems, for the sake of definiteness we focus on the case
of the Edwards-Anderson spin-glass model with Hamiltonian \cite{edwards:75a}
\begin{equation}
  \label{eq:hamiltonian}
  \mathcal{H} = -\sum_{\langle i,j\rangle} J_{ij} s_i s_j - H \sum_i s_i,
\end{equation}
where $s_i = \pm 1$ are Ising spins on a $d$-dimensional lattice chosen in the
present work to be square or simple cubic, applying periodic boundary conditions.
The couplings $J_{ij}$ are quenched random variables which for the examples discussed
here are drawn from either a standard normal distribution or from the discrete
bimodal,
\begin{equation}
  P(J_{ij}) = p\delta(J_{ij}-1) + (1-p)\delta(J_{ij}+1).
  \label{eq:bimodal}
\end{equation}
In zero field, the system undergoes a continuous spin-glass transition in three
dimensions \cite{kawashima:96,hasenbusch:08}, while there is compelling evidence for
a lack of spin-glass order in two-dimensional systems \cite{bhatt:88}.

Two of the basic quantities we consider are the internal energy per spin,
\begin{equation}
  e = \frac{1}{N}[\langle\mathcal{H}(\{s_i\})\rangle],
\end{equation}
where $\langle\cdot\rangle$ denotes a thermal and $[\cdot]$ the disorder average, as
well as the Parisi overlap parameter \cite{parisi:83,young:83},
\begin{equation}
  q = \left[\left\langle\frac{1}{N}\sum_i s_i^{(1)}s_i^{(2)}\right\rangle\right],
  \label{eq:overlap}
\end{equation}
which takes a non-zero value in the spin-glass (but also in a ferromagnetic
\cite{berg:02a}) phase.  Here, $s_i^{(1)}$ and $s_i^{(2)}$ denote the spins of two
independent systems with the same disorder configuration but different stochastic
time evolutions simulated in parallel.

\subsection{Parallel tempering simulations}

\begin{figure}[!tb]
  \centering
  \includegraphics[scale=0.4]{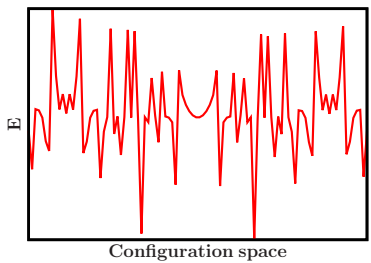}
  \caption {Schematic representation of the rugged energy landscape of a system with
    strong disorder such as a spin glass.}
  \label{fig:rugged}
\end{figure}

While for simpler systems single-spin flip simulations using, for instance, the
transition probabilities proposed by Metropolis {\em et al.} \cite{metropolis:53a}
are sufficient to approach stationarity of the Markov chain, this is much more
difficult for problems with strong disorder exhibiting a rugged (free) energy
landscape as schematically depicted in Fig.~\ref{fig:rugged}. At temperatures where
the typical energy is below that of the highest barriers, simulations with local,
canonical dynamics are not able to explore the full configuration space at reasonable
time scales and instead get trapped in certain valleys of the (free) energy
landscape. The parallel tempering approach \cite{hukushima:96a} attempts to alleviate
this problem by the parallel simulation of a sequence of replicas of the system
running at different temperatures. Through equilibrium swap moves of replicas usually
proposed for adjacent temperature points, copies that are trapped in one of the
metastable states at low temperatures can travel to higher temperatures where
ergodicity is restored. Continuing their random walk in temperature space, replicas
ultimately wander back and forth between high and low temperatures and thus explore
the different valleys of the landscape according to their equilibrium weights. More
precisely, the approach involves simulating $N_T$ replicas at temperatures
$T_0 < T_1 < \cdots < T_{N_T-1}$. As is easily seen, in order to satisfy detailed
balance the proposed swap of two replicas running at temperatures $T_i$ and $T_j$
should be accepted with probability \cite{hukushima:96a}
\begin{equation}
  \label{eq:PT-probability}
  p_\mathrm{acc} = \min\left[1,e^{(1/T_i-1/T_j)(E_i-E_j)}\right],
\end{equation}
where $E_i$ and $E_j$ denote the configurational energies of the two system
replicas. If the swap is accepted, replica $i$ will now evolve at temperature $T_j$
and replica $j$ at temperature $T_i$. On the technical side, it is clear that this
can be seen, alternatively, as an exchange of spin configurations or as an exchange
of temperatures. While the method does not provide a magic bullet for hard
optimization problems where the low lying states are extremely hard to find (such as
for``golf course'' type of energy landscapes) \cite{machta:09}, it leads to
tremendous speed-ups in simulations of spin glasses in the vicinity and below the
glass transition, and it has hence established itself as the de facto standard
simulational approach for this class of problems.

\subsection{Choice of temperature set}

The parallel tempering scheme exhibits a number of adjustable parameters which can be
tuned to achieve acceptable or even optimal performance. These include, in
particular, the temperature set $\{T_i\}$ as well as the set $\{\theta_i\}$ of the
numbers of sweeps of spin flips to be performed at temperature $i$ before attempting
a replica exchange move. In the majority of applications $\theta_i = \theta$ is
chosen independent of the temperature point, and there is some indication that more
frequent swap proposals in general lead to better mixing, such that swaps are often
proposed after each sweep of spin flips (for some theoretical arguments underpinning
this choice see Ref.~\cite{dupuis:12}). Simple schemes for setting the temperature
schedule that have been often employed use certain fixed sequences such as an
inversely linear temperature schedule, corresponding to constant steps in inverse
temperature $\beta$, or a geometric sequence where temperatures increase by a
constant factor at each step \cite{predescu:04,rozada:19}. For certain problems these
work surprisingly well, but there is no good way of knowing a priori whether this
will be the case for a given system.

\begin{figure}[tb!]
  \centering
  \includegraphics[scale=0.4]{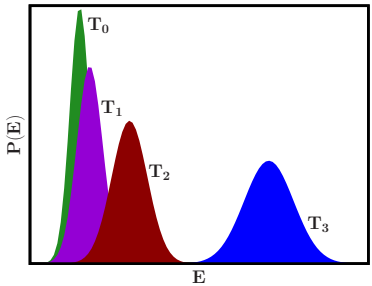}
  \caption{Schematic illustration of the canonical energy histograms at different
    temperatures. For parallel tempering to work efficiently, all pairs of
    neighboring temperatures need to have sufficient histogram overlap, such as for
    the temperatures $T_0$, $T_1$ and $T_2$ shown here. Temperature $T_3$, on the
    other hand, does not provide sufficient overlap with the simulation at $T_2$,
    leading to a poor acceptance rate for the exchange moves.}
  \label{fig:histograms}
\end{figure}

In order to minimize bias (systematic error) and statistical error, it is clear that
the optimal temperature and sweep schedules will result in minimal relaxation times
into equilibrium and decorrelation times in equilibrium. As these times are
relatively hard to accurately determine numerically \cite{sokal:97}, and they also
depend on the observables considered, it is convenient to instead focus on the
minimization of the round-trip or {\em tunneling time\/} $\tau_\mathrm{tunnel}$,
i.e., the expected time it takes for a replica to move from the lowest to the highest
temperature and back, which is a convenient proxy for the more general spectrum of
decorrelation times. The authors of Ref.~\cite{katzgraber:06} proposed a method for
rather directly minimizing $\tau_\mathrm{tunnel}$ by placing temperatures in a way
that maximizes the local diffusivity of replicas, but the technique is rather
elaborate and the numerical differentiation involved can make it difficult to
control. As the random walk of replicas in temperature space immediately depends on
the replica exchange events, it is a natural goal to ensure that such swaps occur
with a sufficient probability at all temperatures \cite{predescu:04}. It is clear
from Eq.~\eqref{eq:PT-probability} that these probabilities correspond to the overlap
of the energy histograms at the adjacent temperature points, see also the
illustration in Fig.~\ref{fig:histograms}. Different approaches have been used to
ensure a constant histogram overlap, either by iteratively moving temperature points
\cite{hukushima:99}, or by pre-runs and the use of histogram reweighting
\cite{bittner:08}. Interestingly, however, constant overlaps do not, in general, lead
to minimum tunneling times in cases where the autocorrelation times of the employed
microscopic dynamics have a strong temperature dependence, but optimal tunneling can
be achieved when using constant overlap together with a sweep schedule taking the
temperature dependence of autocorrelation times into account, i.e., by using
$\theta_i \sim \tau_\mathrm{can}(T_i)$ \cite{bittner:08}.

\begin{figure}[tb!]
  \centering
  \includegraphics[scale=0.5]{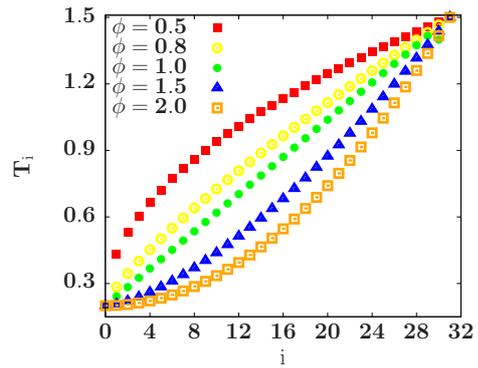}
  \caption{Temperature sequence of the family defined through
    Eqs.~\eqref{eq:protocol} and \eqref{eq:protocol2} for $T_\mathrm{min} = 0.2$ and
    $T_\mathrm{max} = 1.5$ with $N_T = 32$ temperature points and different values of
    the exponent $\phi$.  }
  \label{fig:spacing}
\end{figure}

While the various optimized schemes lead to significantly improved performance, the
additional computational resources required for the optimization are rather
substantial, in particular for disordered systems where the optimization needs to be
performed separately for each disorder sample. As an alternative we suggest to
directly optimize the tunneling times among a suitably chosen family of temperature
schedules. The corresponding family of temperature sequences is given by
\begin{equation}
  T_i = i^{\phi} T_{\rm norm} + T_{\rm min},
  \label{eq:protocol}
\end{equation}
where $\phi$ is a free parameter and
\begin{equation}
  T_{\rm norm} = \displaystyle \frac{T_{\rm max}-T_{\rm min}}{(N_T-1)^{\phi}},
  \label{eq:protocol2}
\end{equation}
and $N_T$ as before denotes the number of temperatures. Clearly, the case $\phi=1$
corresponds to a linear schedule, while $\phi < 1$ and $\phi > 1$ result in the
temperature spacing becoming denser towards higher and lower temperatures,
respectively. This is illustrated in Fig.~\ref{fig:spacing}, where we show the
temperature spacing resulting from different choices of $\phi$ while keeping
$T_\mathrm{min}$, $T_\mathrm{max}$ and $N_T$ fixed.

\begin{figure}[tb!]
  \centering
  \includegraphics[scale=0.5]{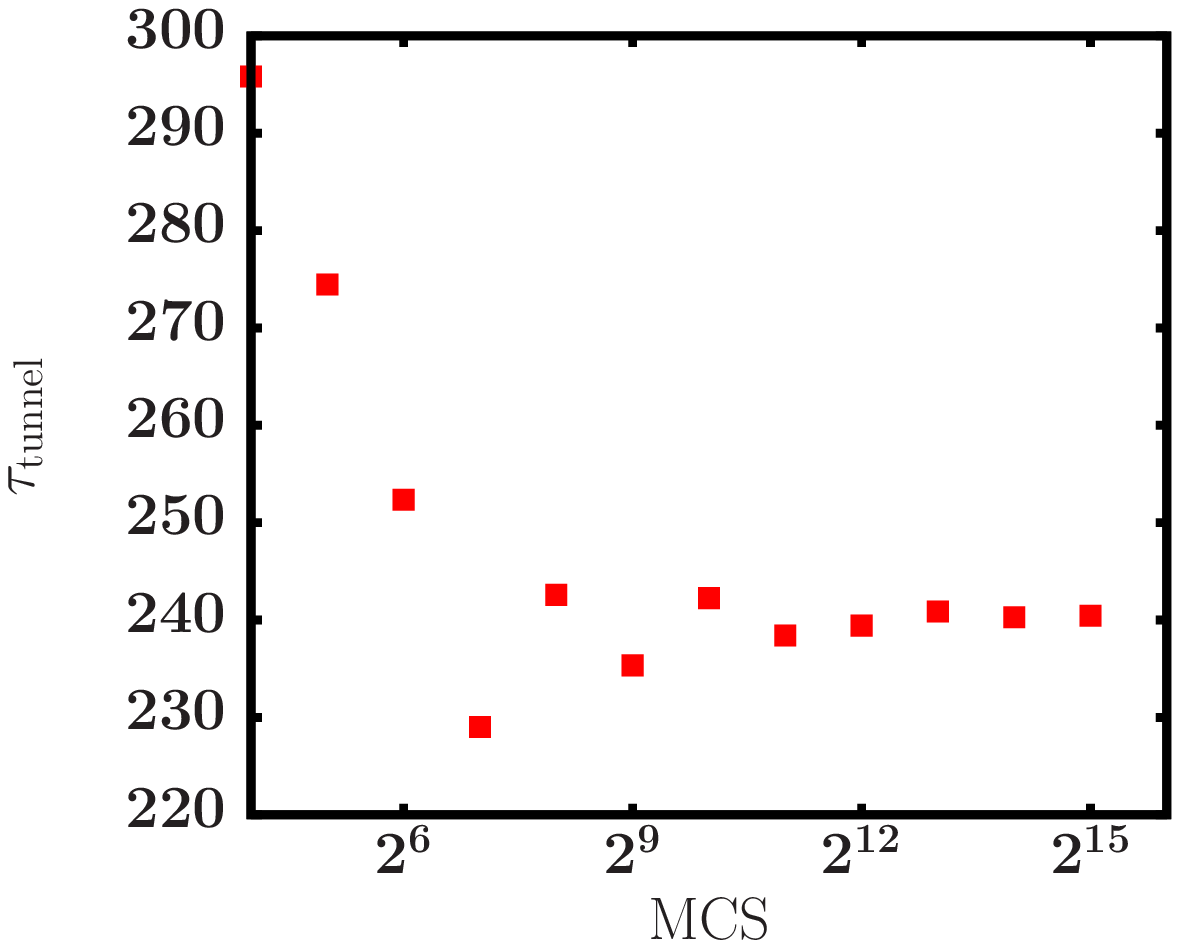}
  \includegraphics[scale=0.5]{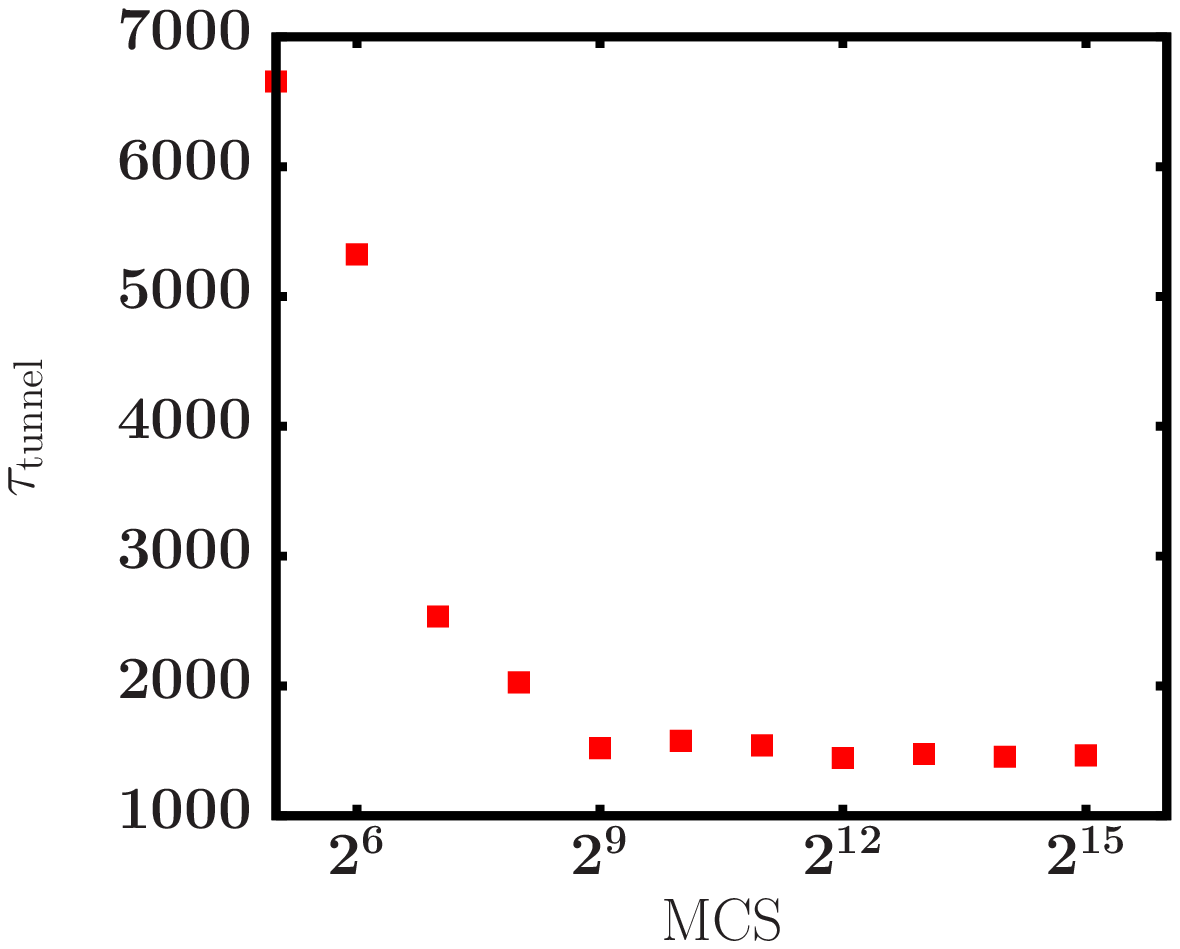}
  \caption{Logarithmically binned time series of tunneling times of parallel
    tempering simulations of the 2D Gaussian Edwards-Anderson model for individual
    samples of size $L=15$ (upper panel) and $L=40$ (lower panel),
    respectively. Here, a Monte Carlo step (MCS) refers to one sweep of spin flips
    for each replica and one (attempted) replica-exchange move for all replicas.}
  \label{fig:tunnel_equilibration}
\end{figure}

\begin{figure}[tb!]
  \centering
  \includegraphics[scale=0.5]{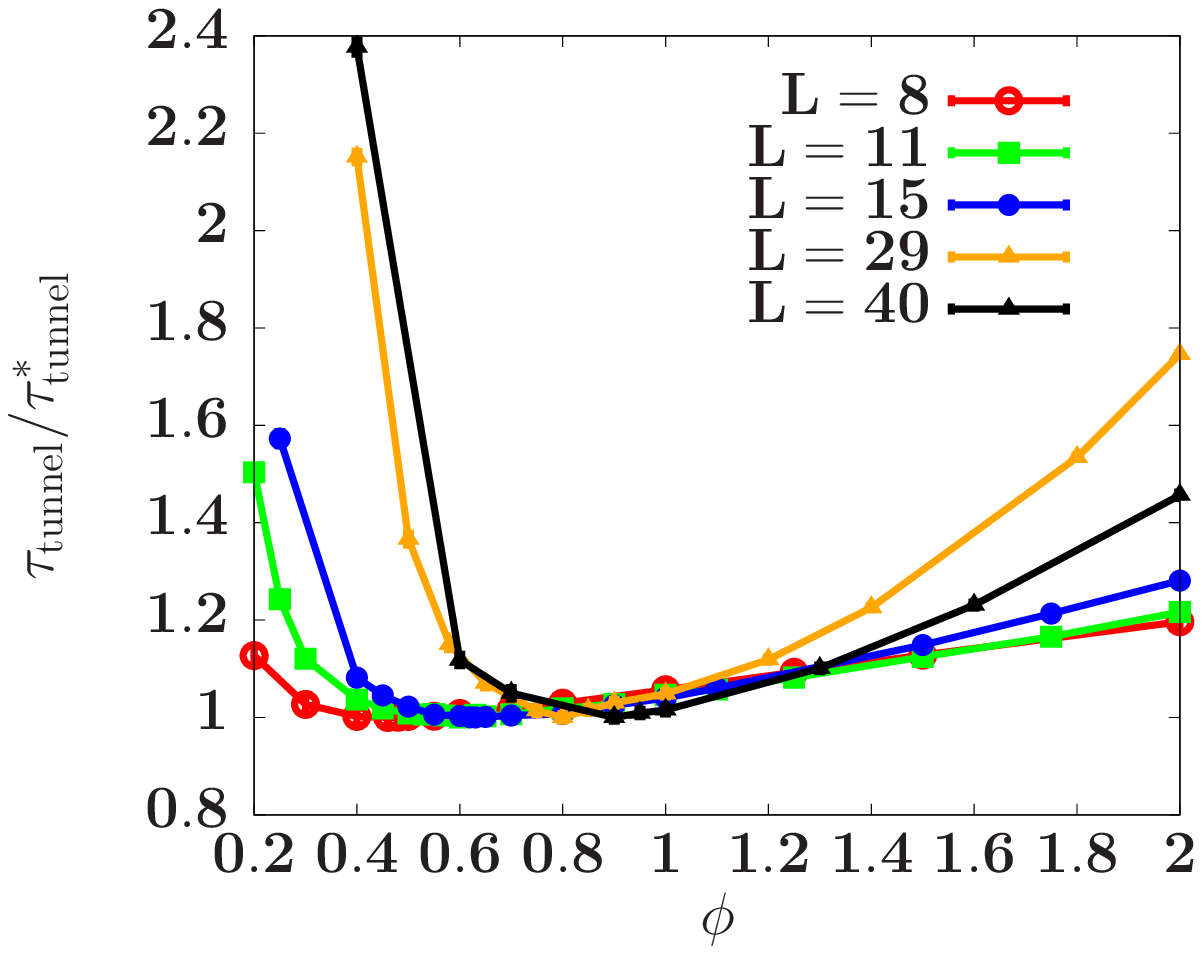}
  \includegraphics[scale=0.5]{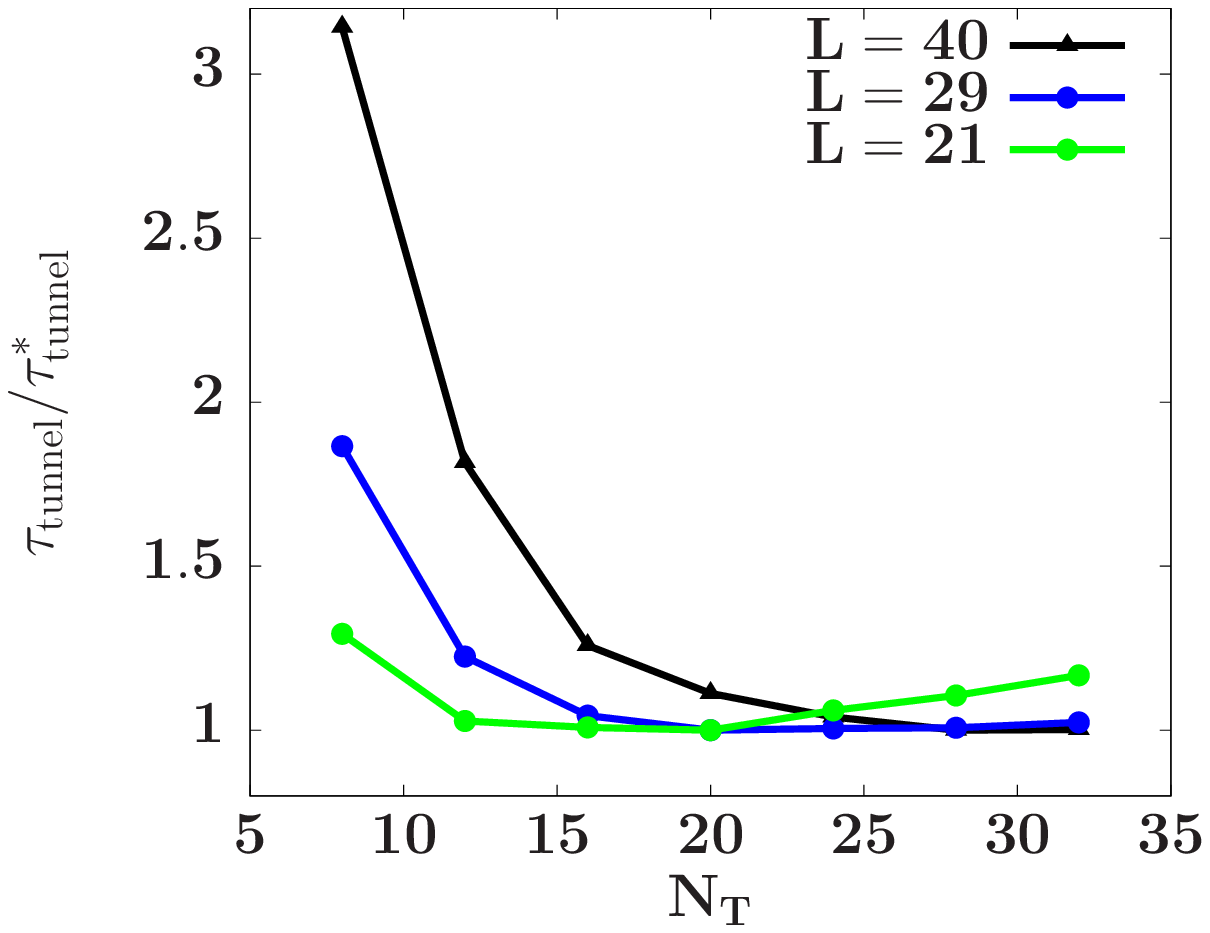}
  \includegraphics[scale=0.5]{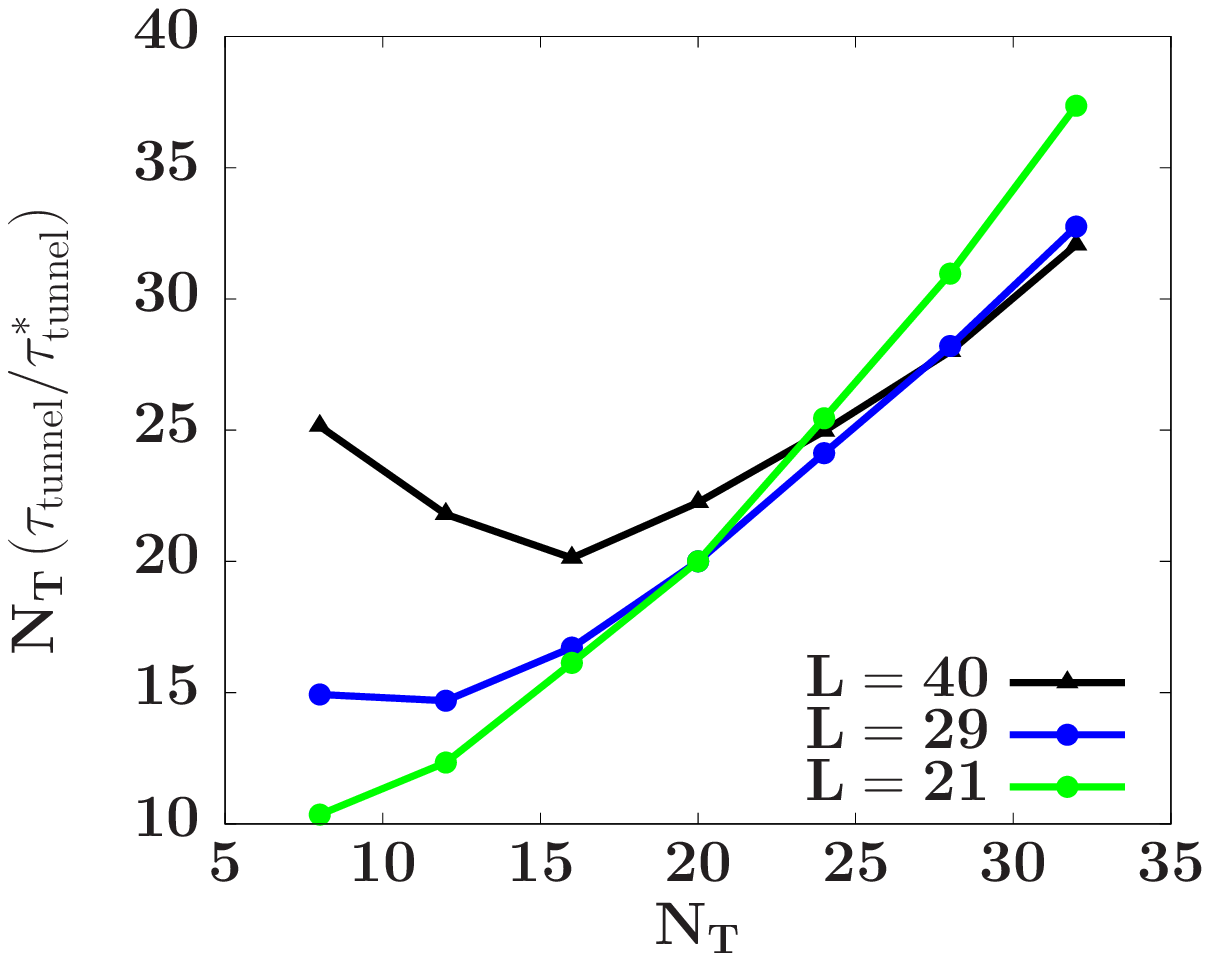}
  \caption[Tunnelling events vs $L/N_T$]{Tunnelling times for parallel tempering
    simulations with temperature protocol \eqref{eq:protocol} relative to their
    minimum $\tau_\mathrm{tunnel}^\ast$ for each system size as a function of the
    exponent $\phi$ for $N_T=32$ (upper panel) and $N_T$ for the optimized value
    $\phi = \phi^\ast$ (lower and bottom panels), respectively.}
  \label{fig:alpha_vs_T}
\end{figure}

To optimize the schedule we vary the parameters in the protocol while monitoring the
resulting tunneling times. In the following we demonstrate this for the case of the
Edwards-Anderson spin glass in two dimensions with Gaussian coupling distribution. We
fix $T_\mathrm{max} = 1.5$, where the system is very quick to relax at any system
size and we choose $T_\mathrm{min} = 0.2$ as the lowest temperature we want to
equilibrate the system at. To arrive at reliable estimates for the tunneling times,
simulations need to be in equilibrium, and we employ the usual logarithmic binning
procedure to ensure this. This is illustrated in Fig.~\ref{fig:tunnel_equilibration}
for systems of size $L=15$ and $L=40$, respectively, where it is seen that the
tunneling times equilibrate relatively quickly as compared to quantities related to
the spin-glass order parameter and so excessively long simulations are not required
for the optimization process. Optimizing the protocol then amounts to choosing the
exponent $\phi$ and the total number $N_T$ of temperature points. To keep the
numerical effort at bay we optimize the two parameters separately. The results of the
corresponding simulations are summarized in Fig.~\ref{fig:alpha_vs_T}. As the top
panel shows, there is a rather clear-cut minimum in the tunneling times as a function
of $\phi$ that shifts from $\phi < 1$ towards larger values of $\phi$ as the system
size $L$ is increased. This trend indicates that for larger systems a higher density
of replicas is required at lower temperatures --- a tendency that is in line with the
general picture of spin-glass behavior.

Likewise it is possible to find the optimal number $N_T$ of temperature points. The
corresponding data from simulations at $\phi = \phi^\ast$ for each system size are
shown in the middle panel of Fig.~\ref{fig:alpha_vs_T}. It is seen that using too few
temperature points leads to rapidly increasing tunneling times. This is a consequence
of too small histogram overlaps in this limit. Too many temperature points, on the
other, are also found to lead to sub-optimal values of $\tau_\mathrm{tunnel}$, which
is an effect of the increasing number of temperature steps that need to be traversed
to travel from the lowest to the highest temperature (and back) as $N_T$ is
increased. This effect only sets in for relatively large $N_T$, however, and as is
seen in the middle panel of Fig.~\ref{fig:alpha_vs_T} there is a rather shallow
minimum for sufficiently large numbers of replicas. To find the optimum investment in
computer time, on the other hand, it is useful to also consider the tunneling time in
units of (scalar) CPU time, which is expected to be proportional to
$N_T\tau_\mathrm{tunnel}$. This is shown in the bottom panel of
Fig.~\ref{fig:alpha_vs_T}, where it becomes clear that the optimal choice in terms of
the total computational effort shifts significantly towards smaller $N_T$. If copies
at different temperatures are simulated in parallel, on the other hand, it is more
appropriate to consider the latency instead of the total work, in which case the
choice suggested by the middle panel of Fig.~\ref{fig:alpha_vs_T} is more relevant.

\begin{figure}[!bt]
  \centering
  \includegraphics[scale=0.5]{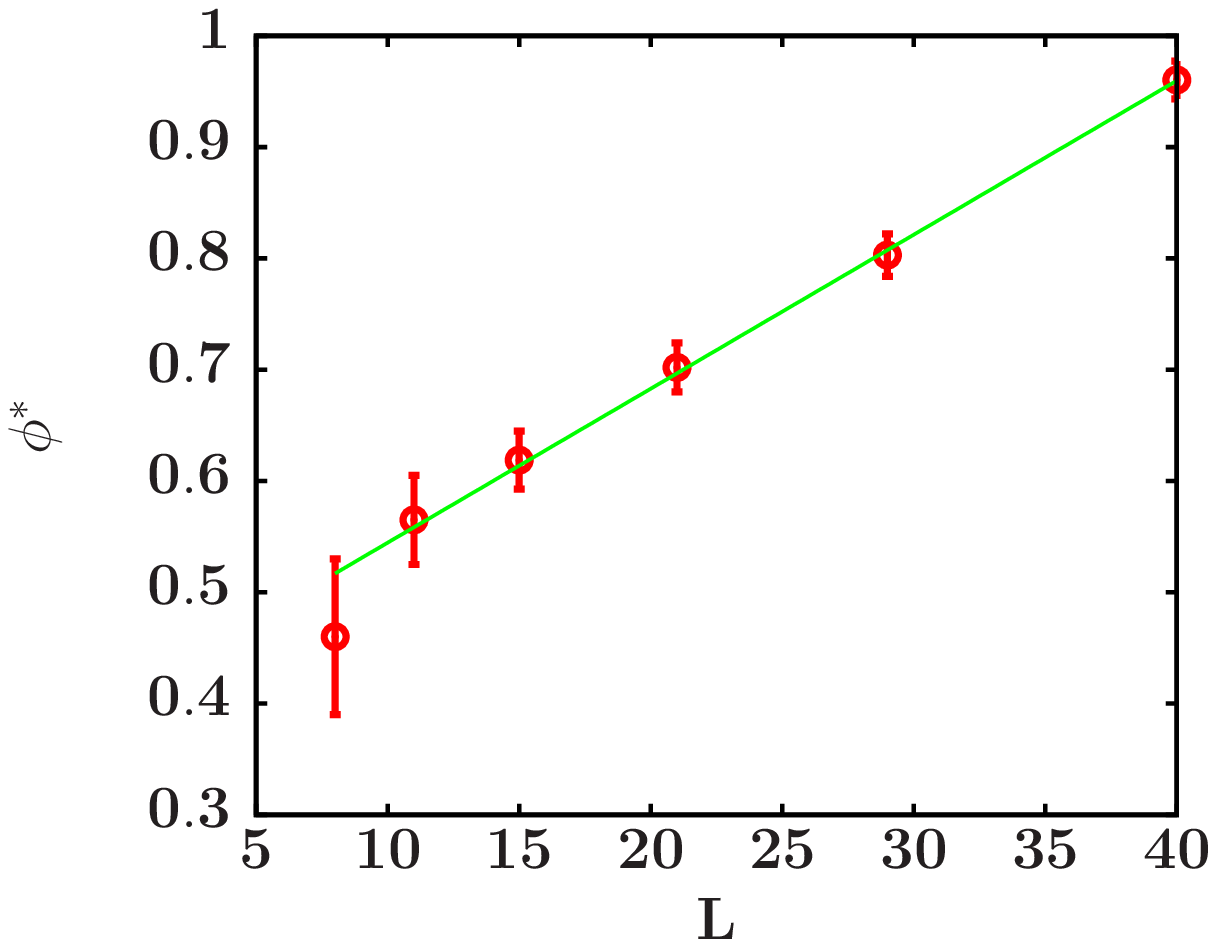}
  \includegraphics[scale=0.5]{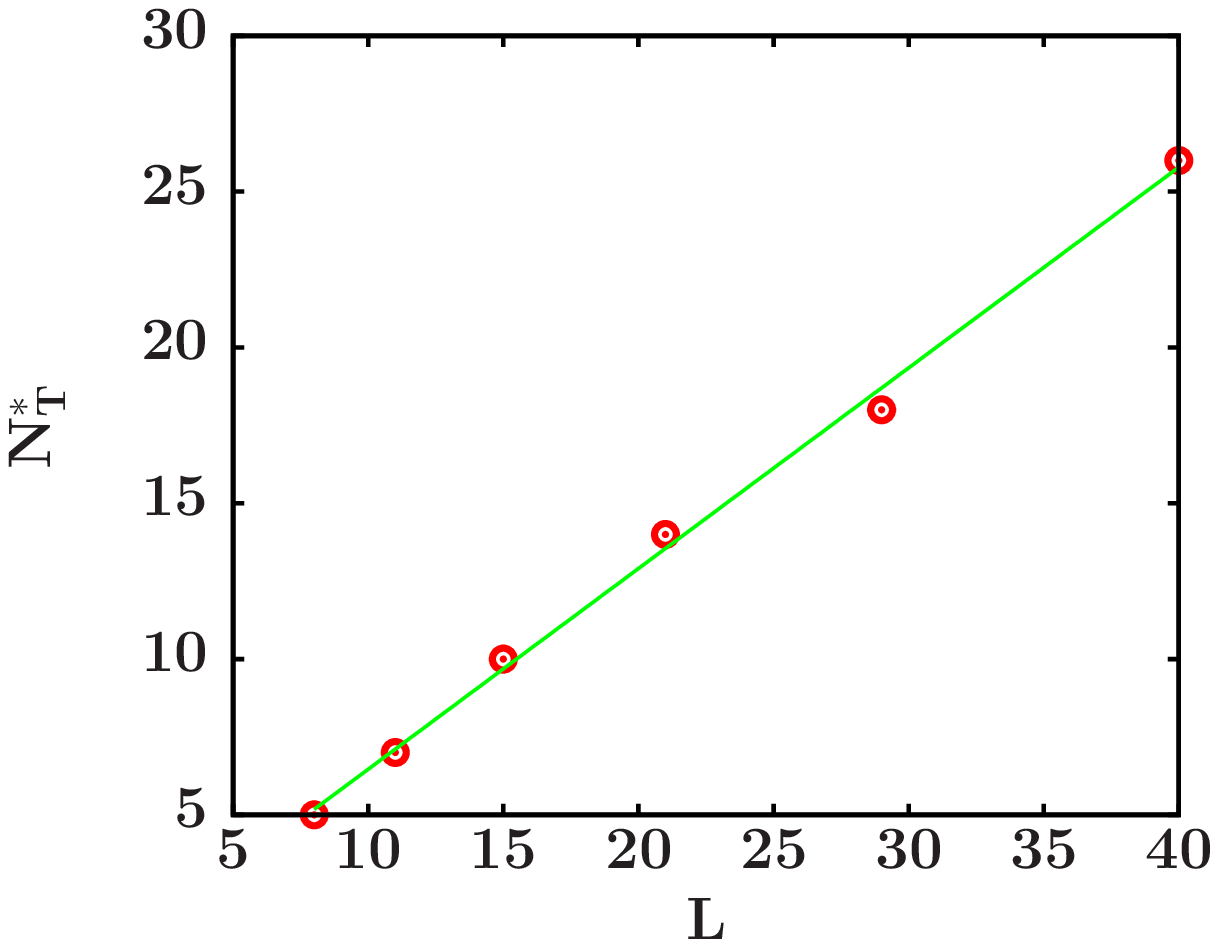}
  \caption{Optimized parameters $\phi^\ast$ and $N_T^\ast$ for parallel tempering
    simulations of the 2D Gaussian Edwards-Anderson spin glass as a function of
    linear system size $L$.}
  \label{fig:L_vs_alpha}
\end{figure}

Good values for $\phi$ and $N_T$ can be inferred for system sizes not directly
simulated by applying finite-size scaling. In Fig.~\ref{fig:L_vs_alpha} we show the
optimal values $\phi^\ast$ and $N_T^\ast$ determined from the procedure above as a
function of system size $L$. As is seen from the plot, both quantities are
approximately linear in $L$, and so we perform fits of the functional form
\begin{equation}
  \label{eq:phi-fit}
  \phi^\ast(L) = a_\phi+b_\phi L
\end{equation}
and
\begin{equation}
  \label{eq:NT-fit}
  N_T^\ast(L) = a_{N_T}+b_{N_T} L
\end{equation}
to the data, resulting in $a_\phi = 0.406$, $b_\phi = 0.014$ and $a_{N_T} = 0.025$,
$b_{N_T} = 0.644$, respectively. We note that on general grounds \cite{janke:08a} we
expect that the required number of temperatures $N_T$ grows like $L^{d/2}$, and so a
steeper than linear increase of $N_T$ is expected in three dimensions. The resulting
fits are convenient mechanisms for predicting good values for the schedule parameters
for larger or intermediate system sizes. We note that both the curves for $\phi$ and
$N_T$ do not show very sharp minima in $\tau_\mathrm{tunnel}$, such that the scheme
is not extremely sensitive with respect to the precise choice of parameter values,
cf.\ the results in Fig.~\ref{fig:alpha_vs_T}.

\subsection{Cluster updates}
\label{sec:cluster-updates}

Even with the help of parallel tempering it remains hard to equilibrate the
spin-glass systems considered here. An additional speed-up of relaxation can come
from the use of non-local updates, in particular for the case of systems in two
dimensions. An efficient cluster update for this case was first proposed by Houdayer
\cite{houdayer:01}. It is similar in spirit to the approach suggested much earlier by
Swendsen and Wang \cite{swendsen:86}, but it uses replicas running at the same
temperature. For two such copies with identical coupling configuration $\{J_{ij}\}$
the method operates on the space of the overlap variables,
\begin{equation}
  \label{eq:local-overlap}
  q_i = s_i^{(1)} s_i^{(2)}.
\end{equation}
cf.\ the definition of the total overlap in Eq.~\eqref{eq:overlap}.  To perform an
update one randomly chooses a lattice site $i_0$ with $q_{i_0} = -1$ and iteratively
identifies all neighboring spins that also have $q_i = -1$, which is most
conveniently done using a breadth-first search. The update then consists in
exchanging the spin configuration of all lattice sites thus identified to belong to
the cluster. As is easily seen, while the energy of both configurations will
potentially change, the total energy $E^{(1)}+E^{(2)}$ remains unaltered. Hence such
moves can always be accepted and the approach is rejection free. For the same reason
it is clearly not ergodic and it hence must be combined with another update such as
single-spin flips to result in a valid Markov chain Monte Carlo method. As it turns
out, the clusters grown in this way percolate (only) as the critical point $T=0$ of
the square-lattice system is approached, and as was demonstrated in
Ref.~\cite{houdayer:01} the method hence leads to a significant speed-up of the
dynamics.

To implement this technique in practice, we hence use the two replicas of the system
at each temperature already introduced in Eq.~\eqref{eq:overlap}. Following
Ref.~\cite{houdayer:01} also larger numbers of replicas at each temperature could be
used, but in practice we find little advantage of such a scheme, and it is instead
more advisable to invest additionally available computational resources into
simulations for additional disorder realisations. We note that with some
modifications the same approach might also be used for systems in three and higher
dimensions \cite{zhu:15a}, although the accelerating effect might be weaker there. In
total our updating scheme hence consists of the following steps:
\begin{enumerate}
\item Perform $N_\mathrm{metro}$ Metropolis sweeps for each replica (usually we
  choose $N_\mathrm{metro}=1$).
\item Perform one Houdayer cluster move for each pair of replicas running with the
  same disorder and at the same temperature.
\item Perform one parallel tempering update for all pairs of replicas running with
  the same disorder at neighboring temperatures.
\end{enumerate}
In the following, we will refer to one such full update as a Monte Carlo step (MCS).

The simulation scheme for spin-glass systems described so far is completely generic,
and it can be implemented with few modifications on a wide range of different
architectures and using different languages. In the following, we will discuss how it
can be efficiently realized on GPUs using CUDA.


\section{Implementation on GPU}
\label{sec:gpu}

From the description of simulation methods in Sec.~\ref{sec:methods} it is apparent
that a simulation campaign for a system with strong disorder naturally leads to a
computational task with manifold opportunities for a parallel implementation:
parallel tempering requires to simulate up to a few dozen replicas at different
temperatures, the measurement of the overlap parameter $q$ and the cluster update
mandate to simulate two copies at each temperature, and finally the disorder average
necessitates to consider many thousands of samples. This problem is hence ideally
suited for the massively parallel environment provided by GPUs. A number of GPU
implementations of spin-glass simulation codes have been discussed previously, see,
e.g., Refs.~\cite{weigel:10a,yavorskii:12,fang:14,baity-jesi:14,lulli:15}. In the
present work we focus on a reasonably simple but still efficient approach that also
allows to include an implementation of the cluster updates that have not previously
been adapted to GPU (but see Ref.~\cite{weigel:10b} for a GPU code for cluster-update
simulations of ferromagnets).

Our implementation targets the Nvidia platform using CUDA \cite{cuda}, but a very
similar strategy would also be successful for other platforms and OpenCL
\cite{scarpino:12}. GPUs offer a hybrid form of parallelism with a number of
multiprocessors that each provide vector operations for a large number of
leightweight threads. Such threads are organized in a grid of blocks that are
independently scheduled for execution \cite{kirk:10,mccool:12}. Of crucial importance
for achieving good performance is the efficient use of the memory hierarchy, and in
particular the goal of ensuring locality of memory accesses of threads in the same
block, as well as the provision of sufficient parallel ``slack'', i.e., the
availability of many more parallel threads than can actively execute instructions on
the given hardware simultaneously \cite{weigel:18}. The latter requirement, which is
a consequence of the approach of ``latency hiding'', where thread groups waiting for
data accesses are set aside in favor of other groups that have completed their loads
or stores and can hence continue execution without delays, is easily satisfied in the
current setup by ensuring that the total number of replicas simulated simultaneously
is sufficiently large.

\begin{figure}[tb!] 
  \centering
  {\includegraphics[scale=.6]{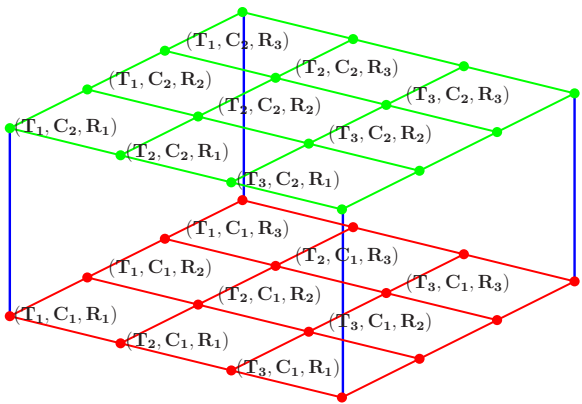}}
  \caption{Distribution of threads in a thread block for the spin updates. Each grid
    point corresponds to a GPU thread, where each thread deals with a copy of the
    spin system for a different combination of temperature $T_i$, replica number
    $C_i$ for cluster updates, and disorder realization $R_i$.}
  \label{fig:thread-block}
\end{figure}

To achieve the goal of locality in memory accesses, also known as ``memory
coalescence'' in the CUDA framework \cite{kirk:10}, it is crucial to tailor the
layout of the spin configurations of different replicas in memory to the intended
access pattern. In contrast to some of the very advanced implementations presented in
Refs.~\cite{yavorskii:12,fang:14,baity-jesi:14,lulli:15}, here we parallelize only
over disorder samples and the replicas for parallel tempering and the cluster update
and hence avoid the use of a domain decomposition and additional tricks such as
multi-spin coding etc. This leads to much simpler code and, as we shall see below, it
still results in quite good performance. Additionally, it has the advantage of being
straightforward to generalize to more advanced situations such as systems with
continuous spins or with long-range interactions. To facilitate the implementation of
the replica-exchange and cluster updates, it is reasonable to schedule the replicas
belonging to the same disorder realization together. We hence use CUDA blocks of
dimension $(N_T, N_C, N_R)$, where $N_T$ is the number of temperatures used in
parallel tempering, $N_C$ is the number of replicas at the same temperature used in
the cluster update, and $N_R$ denotes the number of distinct disorder configurations
simulated in the same block, cf.\ Fig.~\ref{fig:thread-block} for an illustration. In
current CUDA versions, the total number of threads per block cannot exceed $1024$,
and the total number of resident threads per multiprocessor cannot exceed $2048$ (or
$1024$ for compute capability 7.5). It is usually advantageous to maximize the total
number of resident threads per multiprocessor, so a block size of $1024$ threads is
often optimal, unless each thread requires many local variables which then suffer
from spilling from registers to slower types of memory, but this is not the case of
the present problem. To achieve optimal load, it is then convenient to choose $N_T$
as a power of two and, given that we always use $N_C=2$ (cf.\ the discussion in
Sec.~\ref{sec:cluster-updates} above), we then choose $N_R=512/N_T$. It is of course
also possible to use any integer value for $N_T$ and then use
$N_R=\lfloor 512/N_T\rfloor$, leading to somewhat sub-optimal performance. Overall,
we employ $N_B$ blocks, leading to a total of $N_RN_B$ disorder realizations.

In the Metropolis kernel, each thread updates a separate spin configuration, moving
sequentially through the lattice. To ensure memory coalescence, the storage for the
spin configurations is organized such that the spins on the same lattice site but in
different replicas occupy adjacent locations in memory. Note that for each disorder
realization $N_T N_C$ configuration share the sample couplings, such that smaller
overall array dimensions are required for accessing the couplings as compared to
accessing the spins. Random-site spin updates can also be implemented efficiently in
this setup while maintaining memory coalescence by using the same ran\-dom-number
sequence for the site selection but different sequences for the Metropolis criterion
(or other update rule) \cite{gross:17}. Random numbers for the updates are generated
from a sequence of inline generators local to each thread, here implemented in the
form of counter-based Philox generators \cite{salmon:11,manssen:12}. 

To keep things as simple as possible, the parallel tempering update is performed on
CPU \cite{gross:11}. This does not require data transfers as only the configurational
energies are required that need to be transferred from GPU to CPU in any case for
measurements. The actual spin configurations are not transferred or copied as we only
exchange the temperatures. Since even for large-scale simulations no more than a few
dozen temperatures are required, this setup does not create a bottleneck for parallel
scaling. To implement the bidirectional mapping between temperatures and replicas we
use two arrays. On a successful replica exchange move the corresponding entries in
the two arrays are swapped. As well shall see below, this leads to very efficient
code and the overhead of adding the parallel tempering dynamics on top of the spin
flips is quite small.

The cluster update as proposed by Houdayer \cite{houdayer:01} or any generalizations
to higher-dimensional systems \cite{zhu:15a} are implemented in the same general
setup, but now a single thread updates {\em two\/} copies as the cluster algorithm
operates in overlap space. The block configuration is hence changed to
$(N_T, N_C/2, N_R)$. Given that the register usage is not excessive, this decrease
can be compensated by the scheduler by sending twice the number of blocks to each
multiprocessor, such that occupancy remains optimal. Due to the irregular nature of
the cluster growth, however, full coalescence of memory accesses can no longer be
guaranteed, leading to some performance degradation as compared to the spin-flip
kernel. A more profound problem results from the fact that the cluster updates
discussed in Sec.~\ref{sec:cluster-updates} are of the single-cluster nature, such
that differences in cluster sizes lead to deviations in run-time between different
(pairs of) replicas, such that part of the GPU is idling, thus reducing the
computational efficiency. This problem occurs at all levels, from fluctuations
between disorder configurations, to different cluster sizes at different
temperatures, and even fluctuations in the behavior of different pairs of replicas
running with the same disorder and at the same temperature. This is a fundamental
limitation of the approach employed here, and we have not been able to eliminate
it. Particularly important are the variations with temperature as clusters will be
very small at high temperatures and potentially percolating at the lowest
temperatures. Possible steps towards alleviating this effect could be to perform
several cluster updates in the high-temperature copies while waiting for the
low-temperature ones or a formal conversion to a multi-cluster variant which,
however, means that all operations on $q=+1$ clusters (i.e., the exchange of spin
configurations there) leave the systems invariant. These problems notwithstanding,
however, the cluster update if used with moderation where it does not affect the
overall parallel efficiency too strongly is still very useful for speeding up the
equilibration of the system.

Finally, measurements are taken at certain intervals using the same execution
configuration, where measurements of single-replica quantities use blocks of
dimensions $(N_T, N_C, N_R)$ and two-replica quantities such as the spin-glass
susceptibility and the modes for the correlation length use blocks of size
$(N_T, N_C/2, N_R)$.

\begin{figure}[tb!]
  \centering
  \subfigure{\includegraphics[scale=0.43]{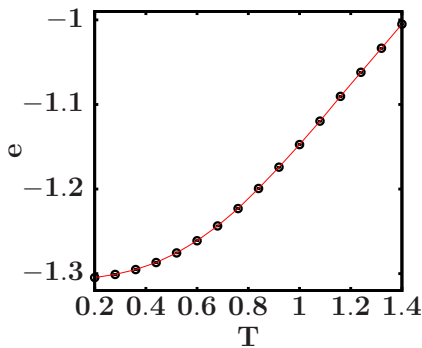}\label{fig:2dgauss}}
  \caption{Average internal energy per spin from parallel tempering, cluster update
    simulations of the 2D Edwards-Anderson spin-glass model with Gaussian couplings
    implemented on GPU. The data for a linear system size $L=15$ are averaged over
    500 disorder samples.  The line shows the exact result as calculated by the
    technique described in Ref.~\cite{galluccio:00}.  }
  \label{fig:results_verification}
\end{figure}

\begin{figure*}[tb!]
  \centering
  \includegraphics[scale=0.35]{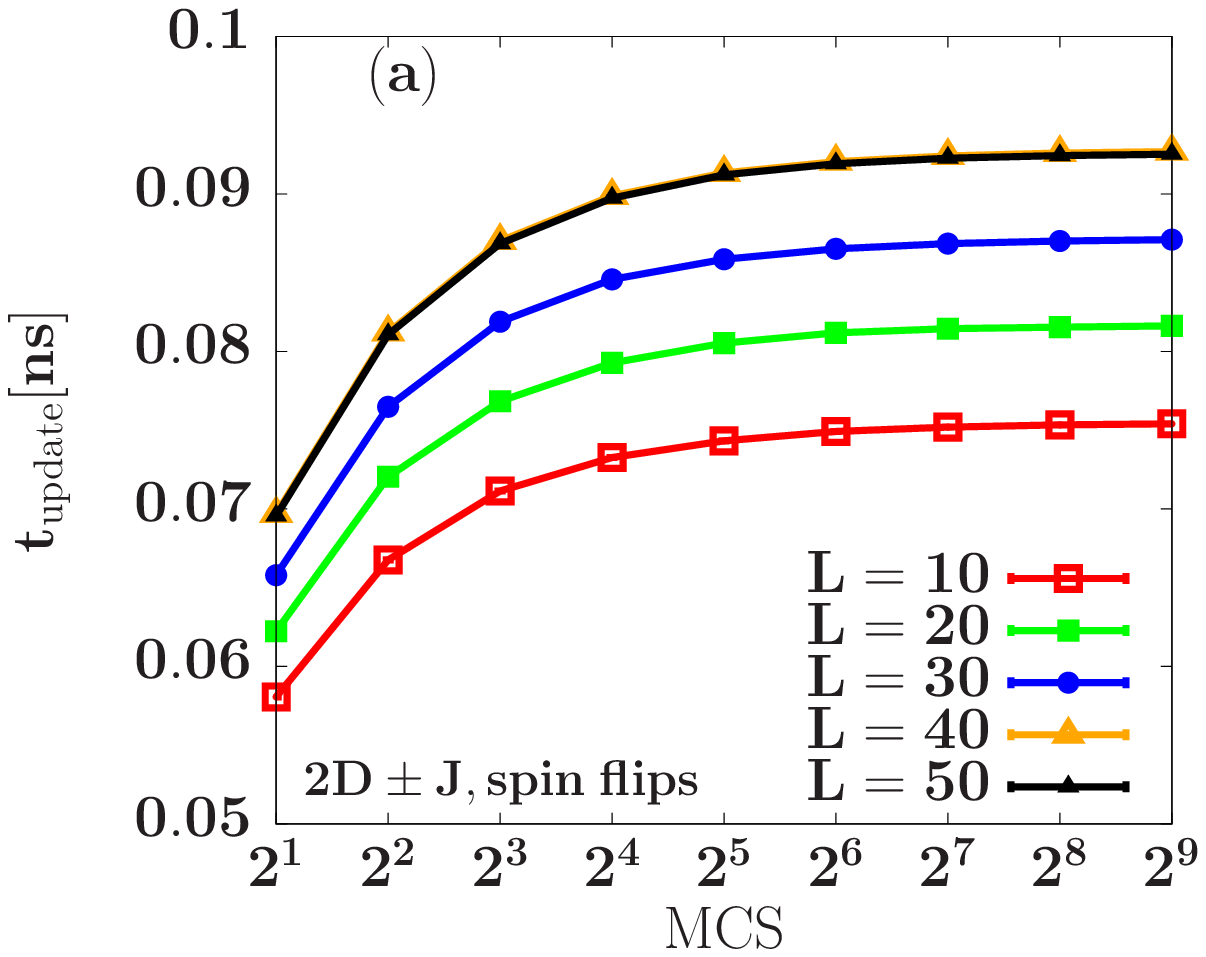}
  \includegraphics[scale=0.35]{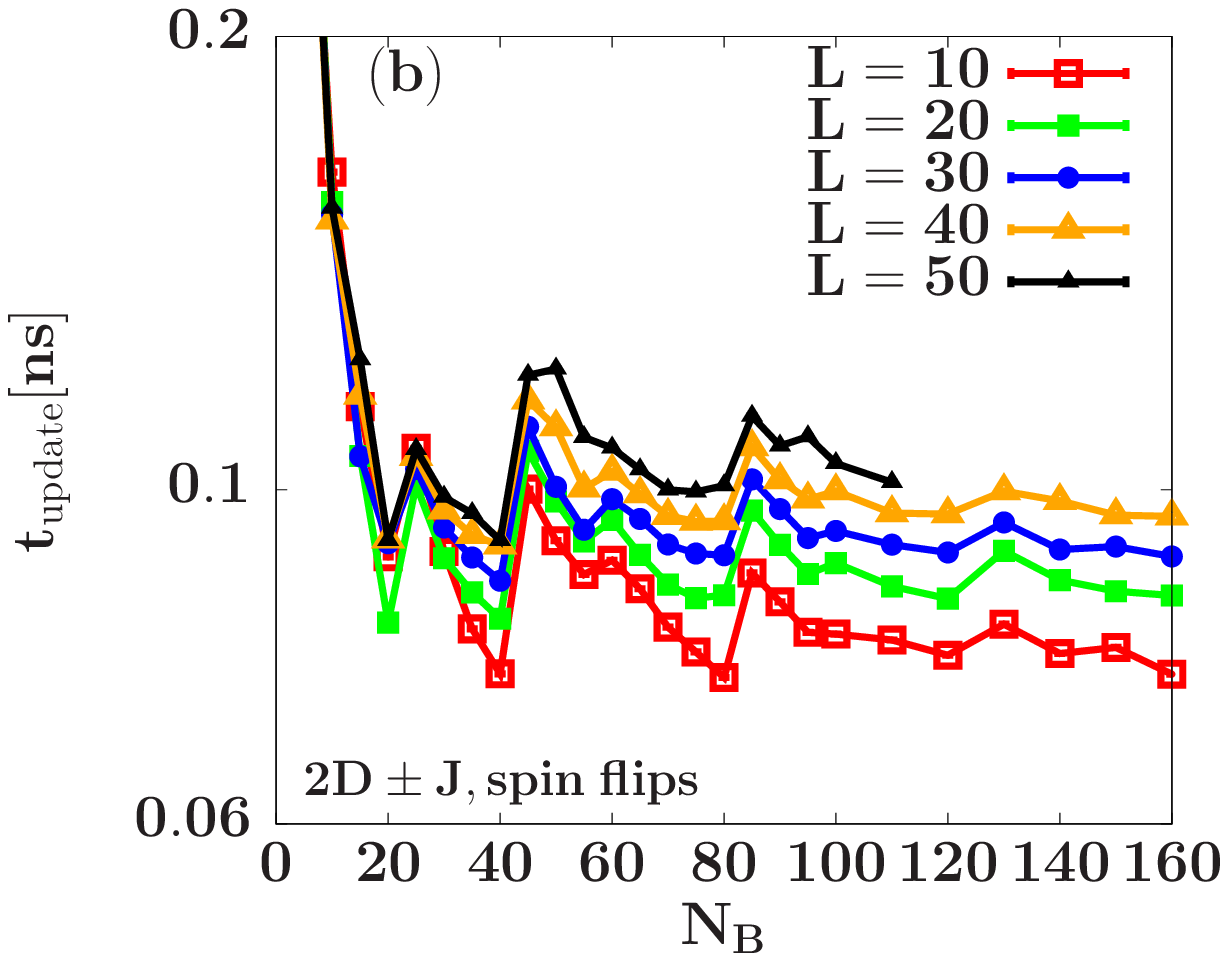}
  \includegraphics[scale=0.35]{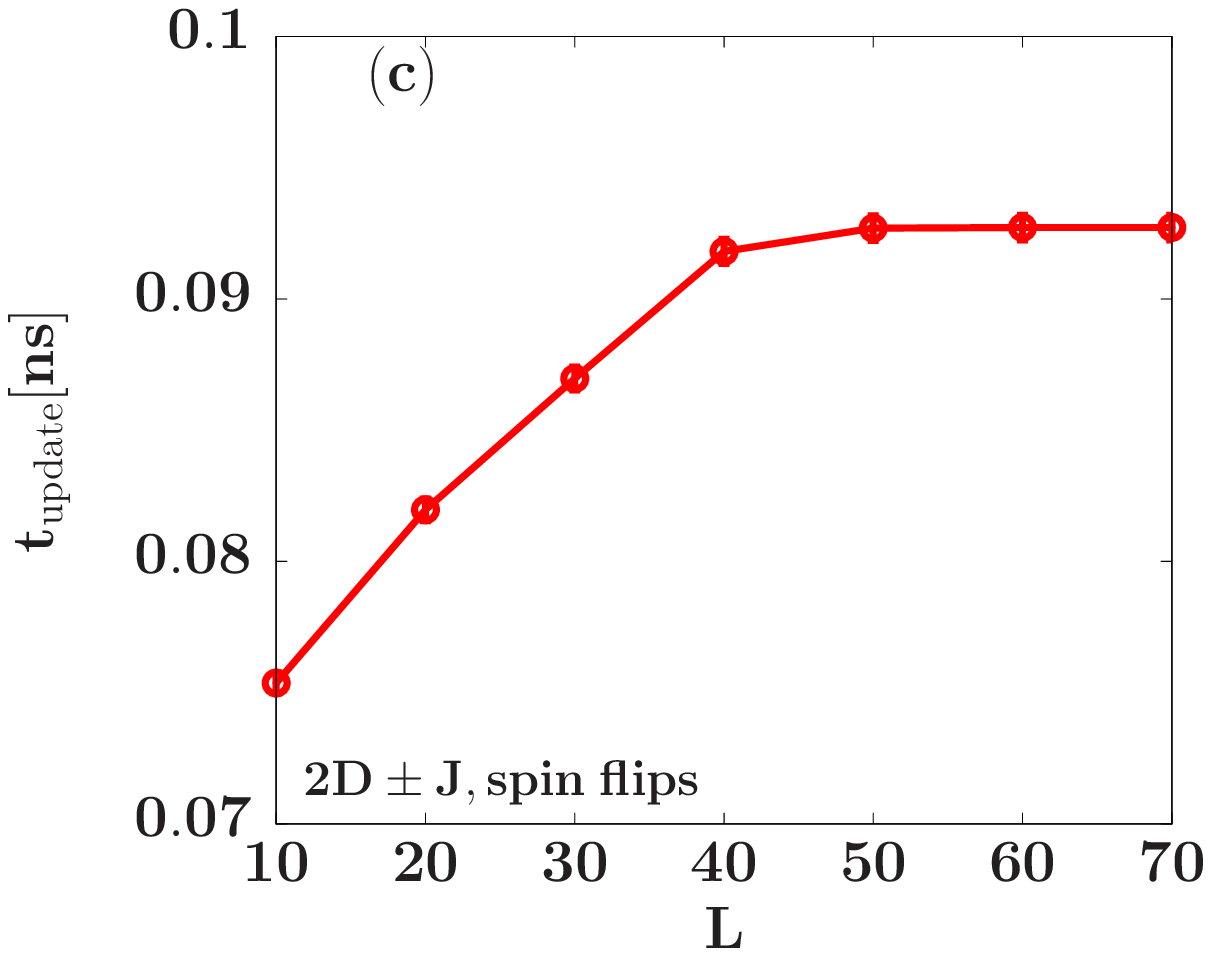}\\
  \includegraphics[scale=0.35]{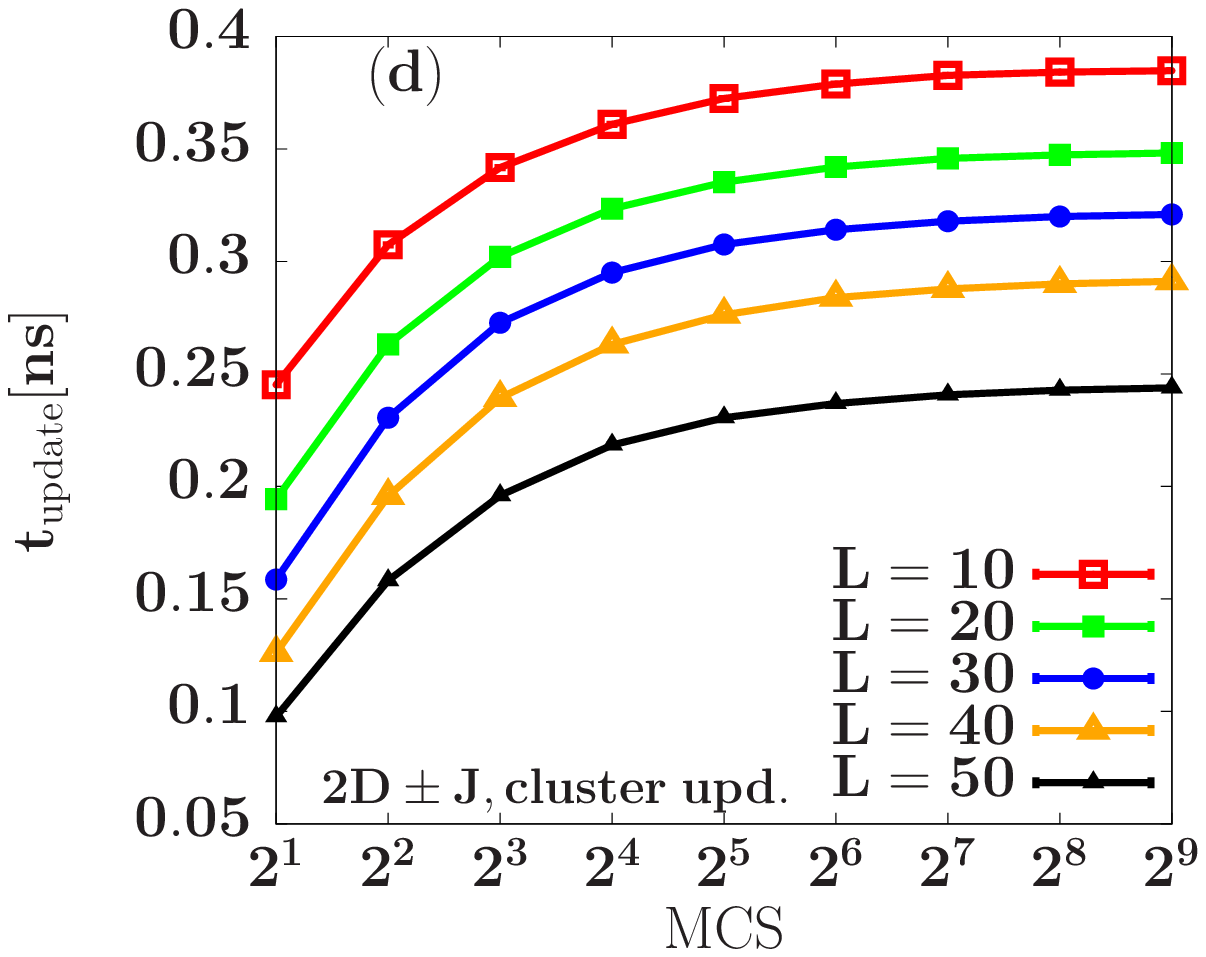}
  \includegraphics[scale=0.35]{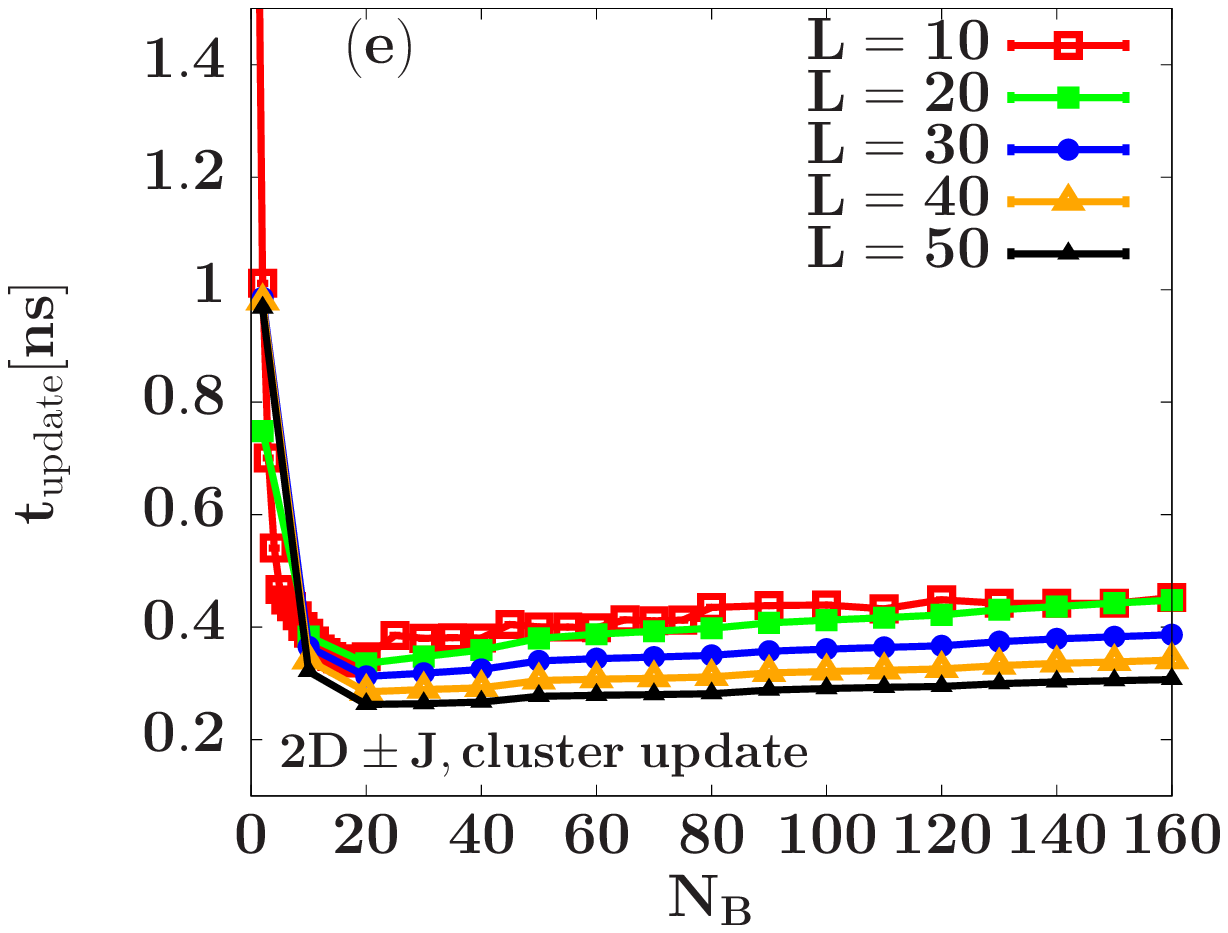}
  \includegraphics[scale=0.35]{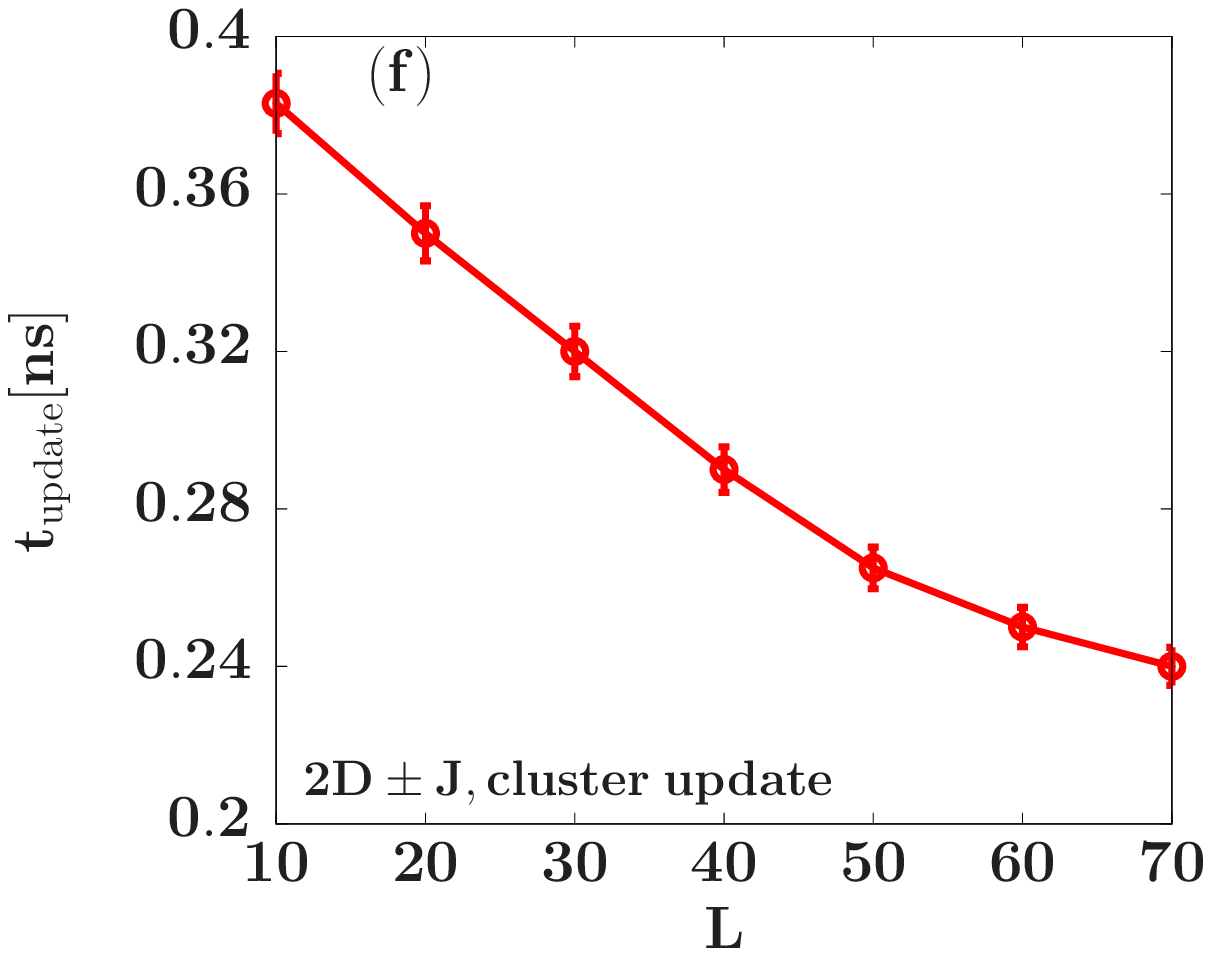}
  \caption{Timing data for simulations of the 2D $\pm J$ Edwards-Anderson model on
    the Nvidia GTX 1080 GPU. (a)--(c) show the time per spin and update spent in the
    Metropolis kernel, while (d)--(f) represent the time spent in the cluster-update
    kernel.}
  \label{fig:gpu_perform}
\end{figure*}

As one step of verification of the correctness of our implementation, we compared the
internal energies as a function of temperature found after careful equilibration to
the exact result found from a Pfaffian technique that allows to determine the
partition function on finite lattices \cite{galluccio:00}. As is apparent from the
comparison performed for a sample of $500$ disorder configurations that is shown in
Fig.~\ref{fig:results_verification} there is excellent agreement, and the simulation
data are compatible with the exact result within error bars.

\section{Performance}
\label{sec:performance}

We assess the performance of the GPU implementation discussed above via a range of
test runs for the Edwards-Anderson model in two and three dimensions, while using
bimodal and Gaussian couplings. For spin-glass simulations, it usually does not make
sense to take very frequent measurements of observables as there are strong
autocorrelations and the fluctuations induced by the random disorder dominate those
that are of thermal origin \cite{picco:98,katzgraber:06a}. Hence the overall run-time
of our simulations is strongly dominated by the time taken to update the
configurations. As is discussed below in Sec.~\ref{sec:tempering-cost}, the time
taken for the parallel tempering moves is small against the time required for spin
flips and the cluster update, and it also does not vary between different models,
such that we first focus on the times spent in the GPU kernels devoted to the
Metropolis and Houdayer cluster updates. To compare different system sizes, we
normalize all times to the number of spins in the system, i.e., we consider
quantities of the form
\begin{equation}
  \label{eq:times}
  t_\mathrm{update} = \frac{t_\mathrm{kernel}}{N_T N_R N_C N_B L^d},
\end{equation}
where $t_\mathrm{kernel}$ is the total GPU time spent in a given kernel. The quantity
$t_\mathrm{update}$ corresponds to the update time per replica and spin for the
considered operation. All benchmarks discussed below were performed on an Nvidia GTX
1080, a Pascal generation GPU with 2560 cores distributed over 20 multiprocessors,
and equipped with 8 GB of RAM. We expect the general trends observed to be
independent of the specific model considered, however.

\subsection{Two-dimensional system with bimodal couplings}

For the case of a bimodal coupling distribution, $J_{ij} = \pm J$, we store the
couplings in 8-bit wide integer variables. For our benchmarks we choose the symmetric
case, i.e., $p=1/2$ in Eq.~\eqref{eq:bimodal}. Considering the time spent in the
Metropolis and cluster-update kernels separately, we first study how the spin-update
times evolve as the systems relax towards equilibrium. In Fig.~\ref{fig:gpu_perform}
(a) and (d) we show how the normalized times according to Eq.~\eqref{eq:times}
develop with the number of Monte Carlo steps (MCS) for the Metropolis and
cluster-update kernels, respectively. It is clear that for both updates, the run
times per step converge relatively quickly, such that after approximately $2^7 = 128$
steps the normalized update times are sufficiently close to stationary. We note that
in the present configuration, the normalized times taken for the cluster update are
about 2--5 times larger than those for the single-spin flips.

\begin{figure*}[tb!]
  \centering
  \includegraphics[scale=0.35]{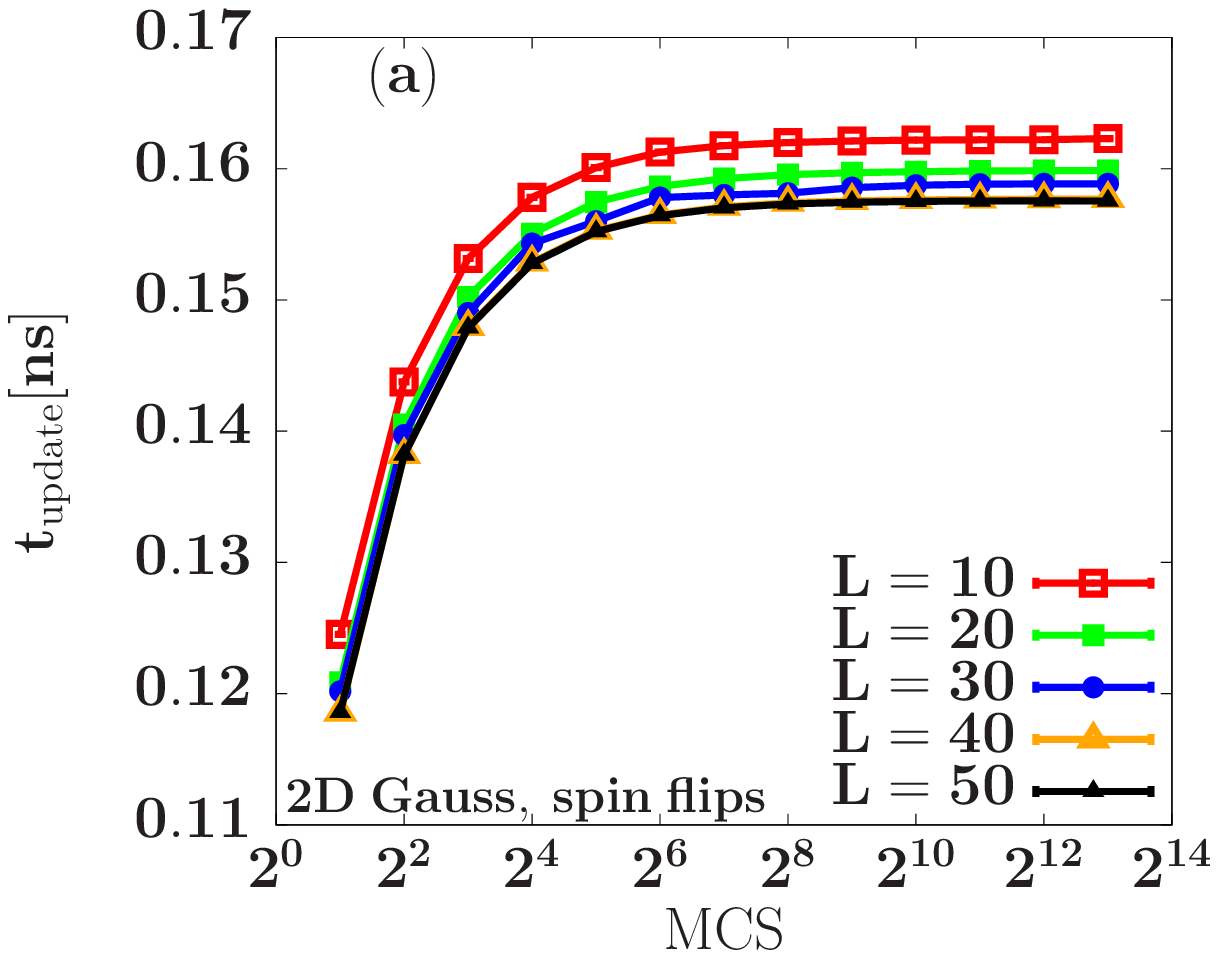}
  \includegraphics[scale=0.35]{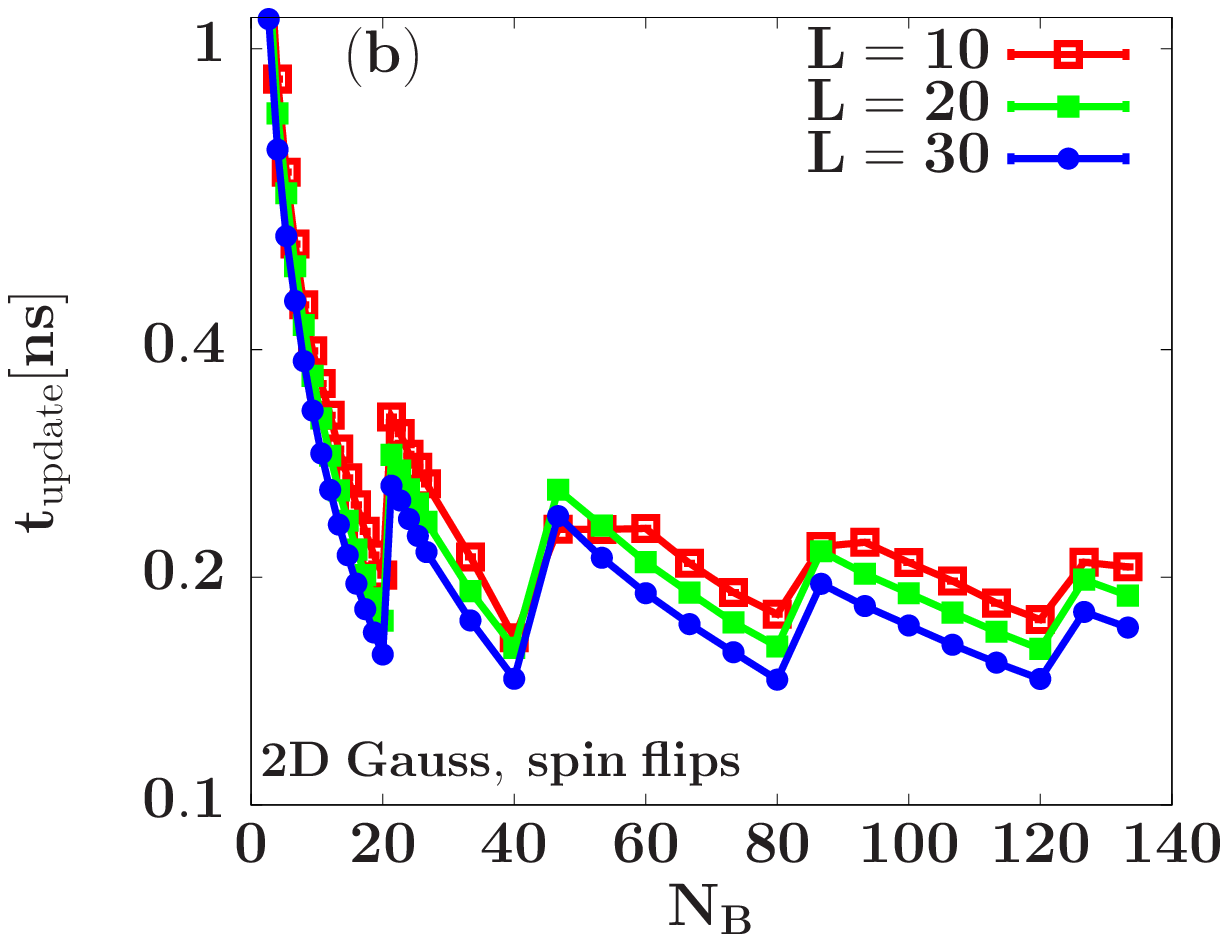}
  \includegraphics[scale=0.35]{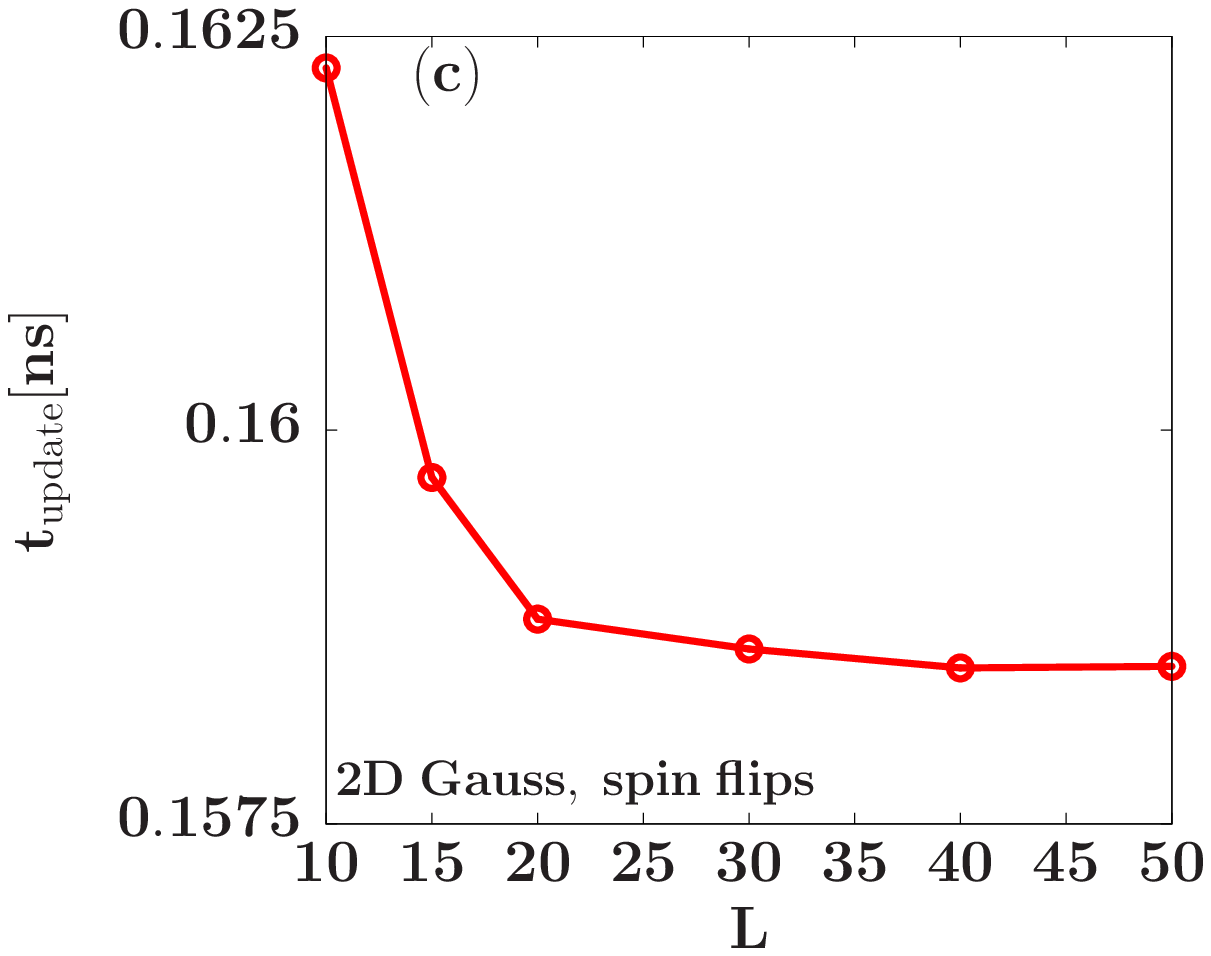}\\
  \includegraphics[scale=0.35]{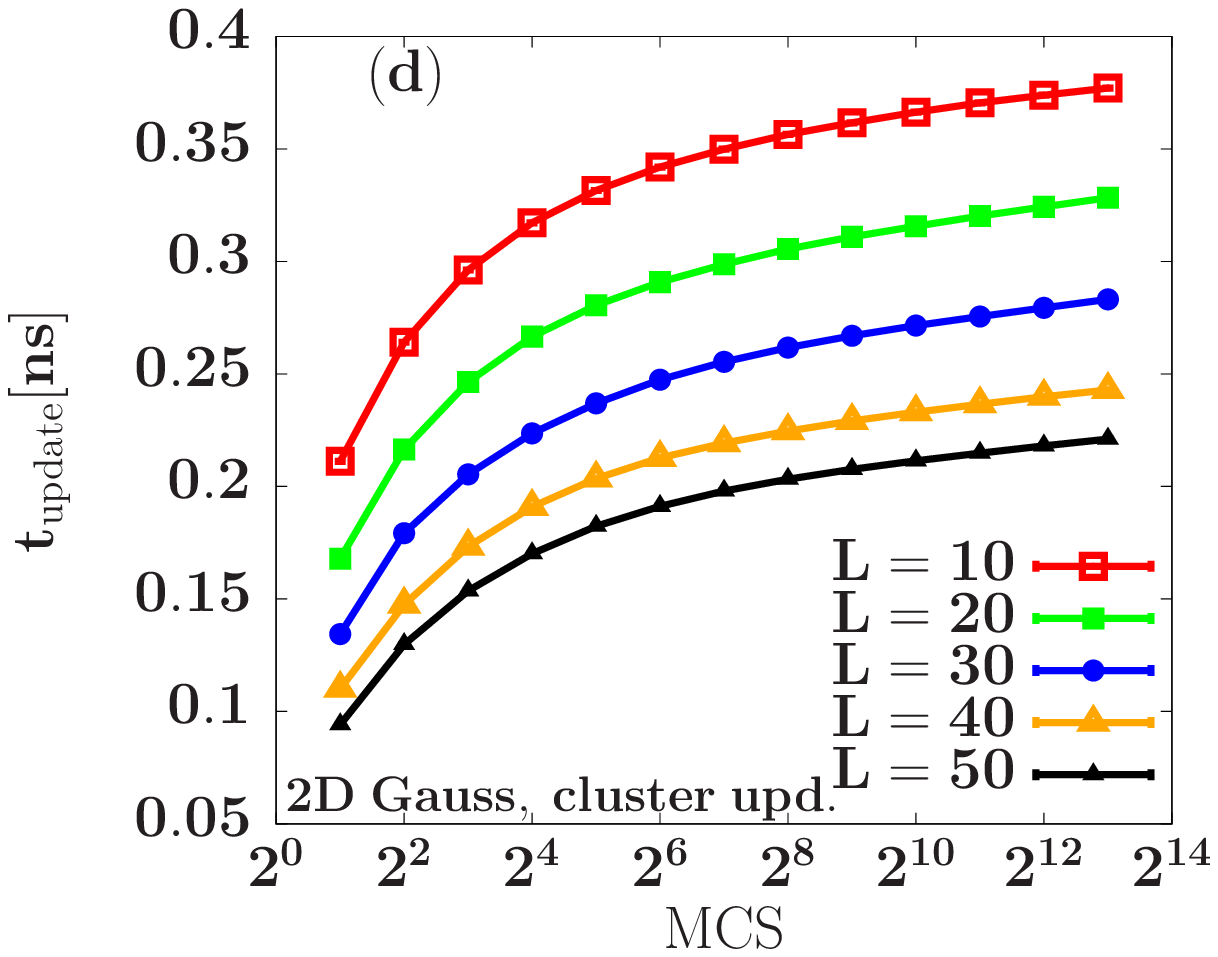}
  \includegraphics[scale=0.35]{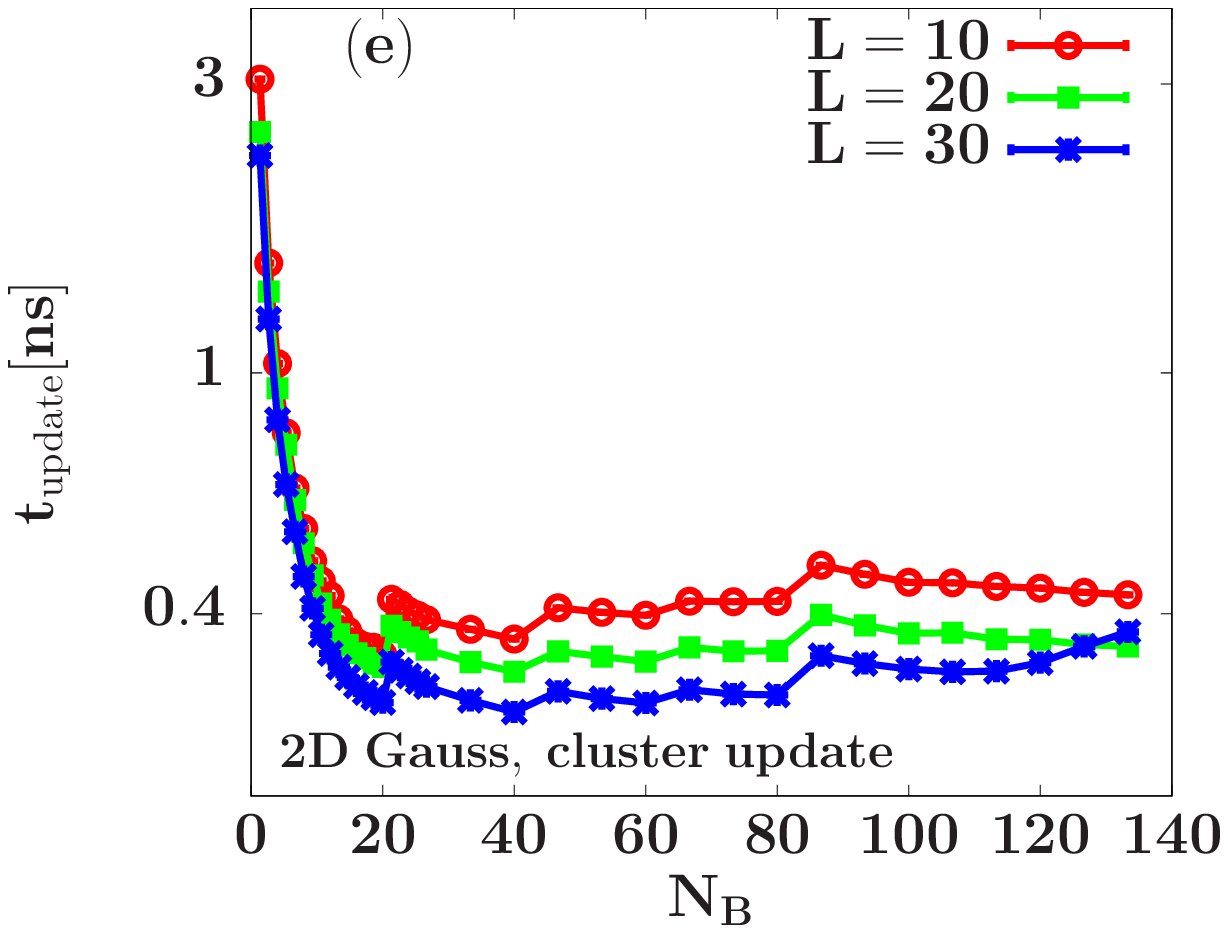}
  \includegraphics[scale=0.35]{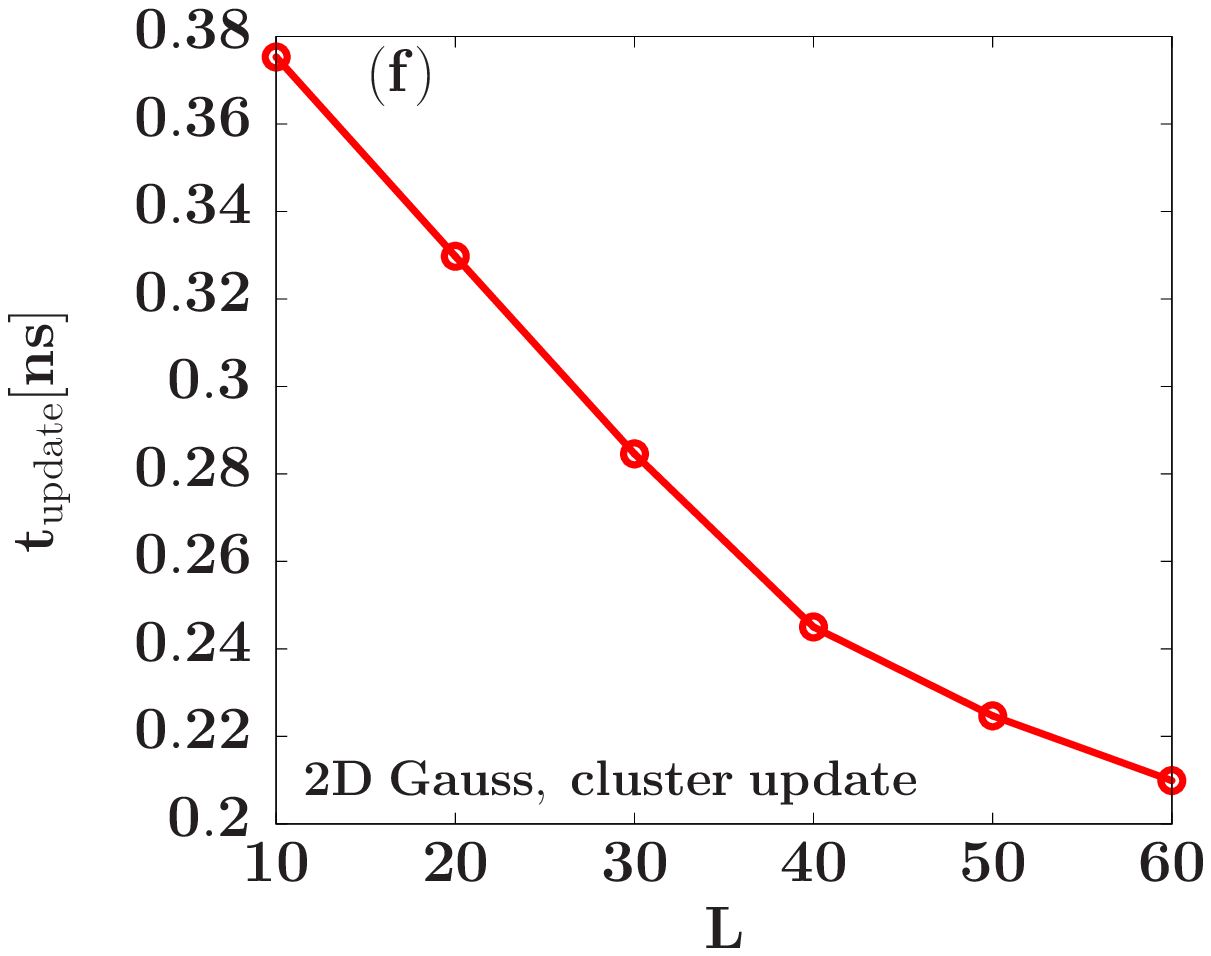}
  \caption{Updating times for GPU simulations of the 2D Gaussian Edwards-Anderson
    model on the GTX 1080 GPU. Panels  (a)--(c) are for the Metropolis kernel, while
    panels (c)--(f) relate to the cluster update.}
  \label{gpu_perform1}
\end{figure*}

As discussed above in Sec.~\ref{sec:gpu}, the size of thread blocks should be
optimized to result in good performance. Often, best results are achieved for cases
of optimal occupancy \cite{weigel:18}, but memory considerations can sometimes shift
the corresponding optima. Occupancy plays the key role for the Metropolis kernel as
is seen in Fig.~\ref{fig:gpu_perform} (b) which shows $t_\mathrm{update}$ as a
function of the number $N_B$ of thread blocks. As discussed above, $N_T$, $N_C$ and
$N_R$ were chosen to result in blocks of $1024$ threads. The limit of $2048$
simultaneously resident threads per multiprocessor means that at most two blocks can
be active on each of the $20$ multiprocessors of the GTX 1080 card. As a consequence,
the best performance is performed for $N_B = 40$ and its multiples, with sub-dominant
minima at multiples of $N_B = 20$. For the cluster update this structure is absent,
cf.\ Fig.~\ref{fig:gpu_perform} (e). This is mostly a consequence of the lack of
memory coalescence resulting from the cluster algorithm: while in the Metropolis
update adjacent threads in a block access adjacent spins in memory, the cluster
construction starts from a random lattice site and proceeds in the form of a
breadth-first search, resulting in strong thread divergence. Our simulations are
hence performed at $N_B = 40$ which is the dominant minimum for the Metropolis kernel
and which is also within the broad minimum for the cluster update.

The lattice-size dependence of GPU performance for the 2D bimodal model is shown in
Fig.~\ref{fig:gpu_perform} (c) and (f). As is seen from Fig.~\ref{fig:gpu_perform}
(c) the smaller system sizes show slightly smaller spin-flip times than the larger
ones. This again is an effect of memory coalescence that is greater for smaller
systems where several rows of spins fit into a single cache line. For the cluster
update, on the other hand, the decrease of spin-flip times with increasing system
size seen in Fig.~\ref{fig:gpu_perform} (f) is an effect of the normalization
according to Eq.~\eqref{eq:times}: in each update only a single cluster is grown and
the cluster size depends on temperature but only weakly on system size, such that the
normalization by $1/L^2$ leads to a decay of $t_\mathrm{update}$ with $L$.

\subsection{Two-dimensional system with Gaussian couplings}

Representative of simulations of systems with a continuous coupling distribution we
consider the case of Gaussian couplings. The code for this case is very similar to
the one for discrete couplings with the difference that now the couplings are stored
in $32$-bit floating-point variables. We again run a range of benchmark
simulations. The results are summarized in Fig.~\ref{gpu_perform1}, showing the
timings for the Metropolis kernel in panels (a)--(c), and those for the cluster
update in panels (d)--(f). Overall, the results are similar to those obtained for the
2D $\pm J$ model, but the updating times are in general somewhat larger for the
Gaussian system since the floating-point operations involved in evaluating the
Metropolis criterion are more expensive than the integer arithmetic required for the
bimodal system. The times for the Metropolis update are almost independent of system
size, clearly showing that the increased coalescence that was responsible for higher
performance on smaller systems in the bimodal system is no longer relevant here as
more time is spent on fetching the couplings and evaluating the acceptance criterion.

\begin{figure*}[tb!]
  \centering
  \includegraphics[scale=0.35]{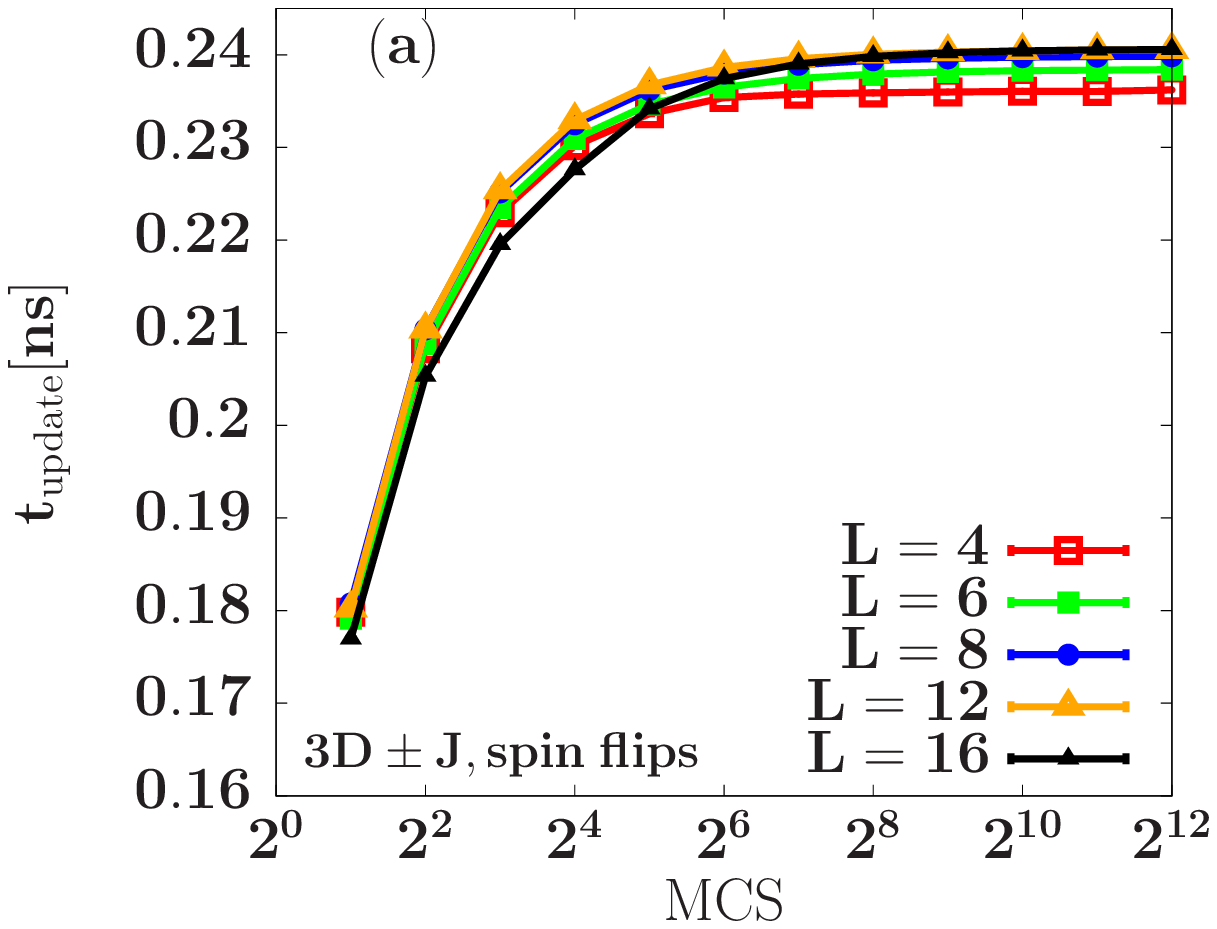}
  \includegraphics[scale=0.35]{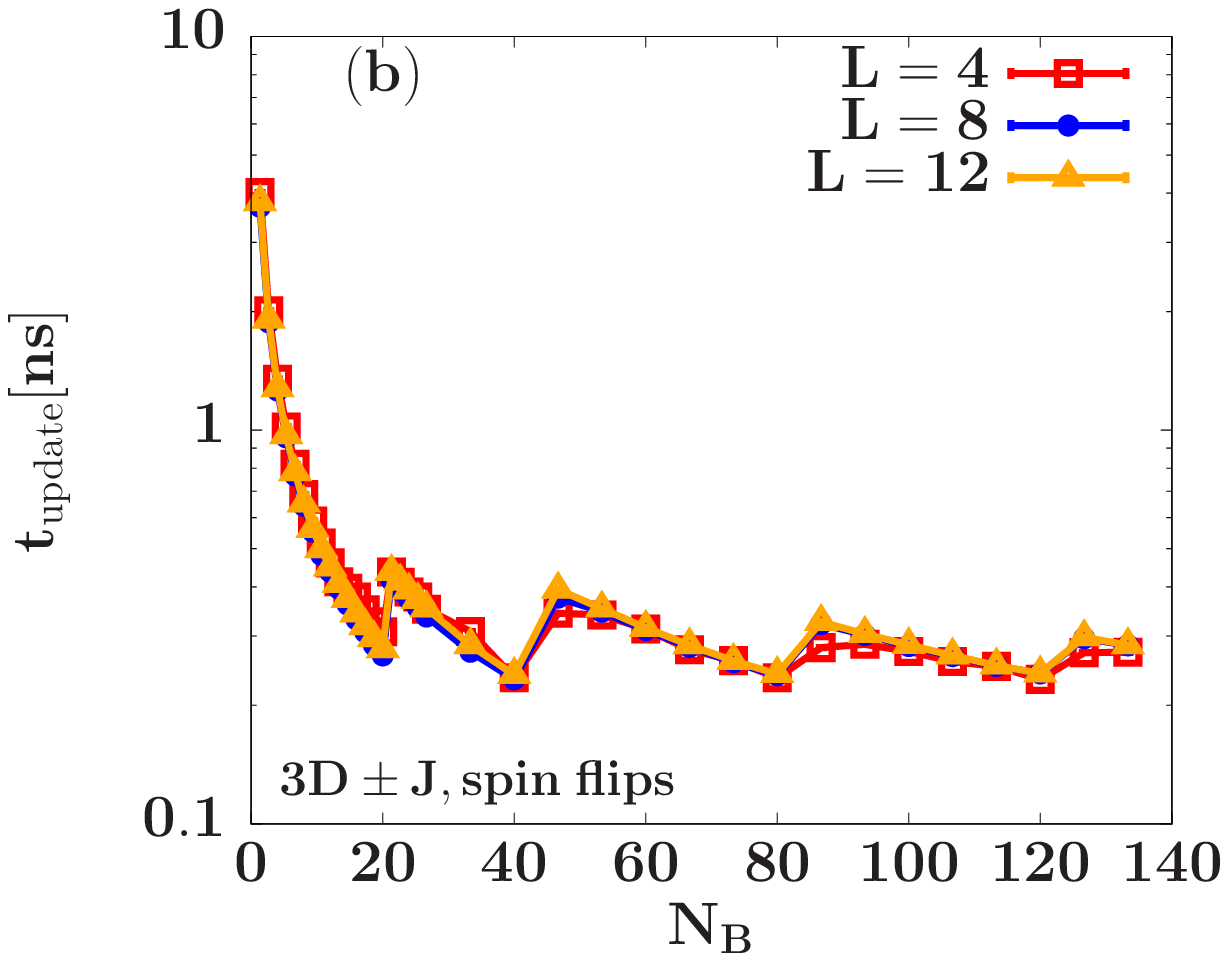}
  \includegraphics[scale=0.35]{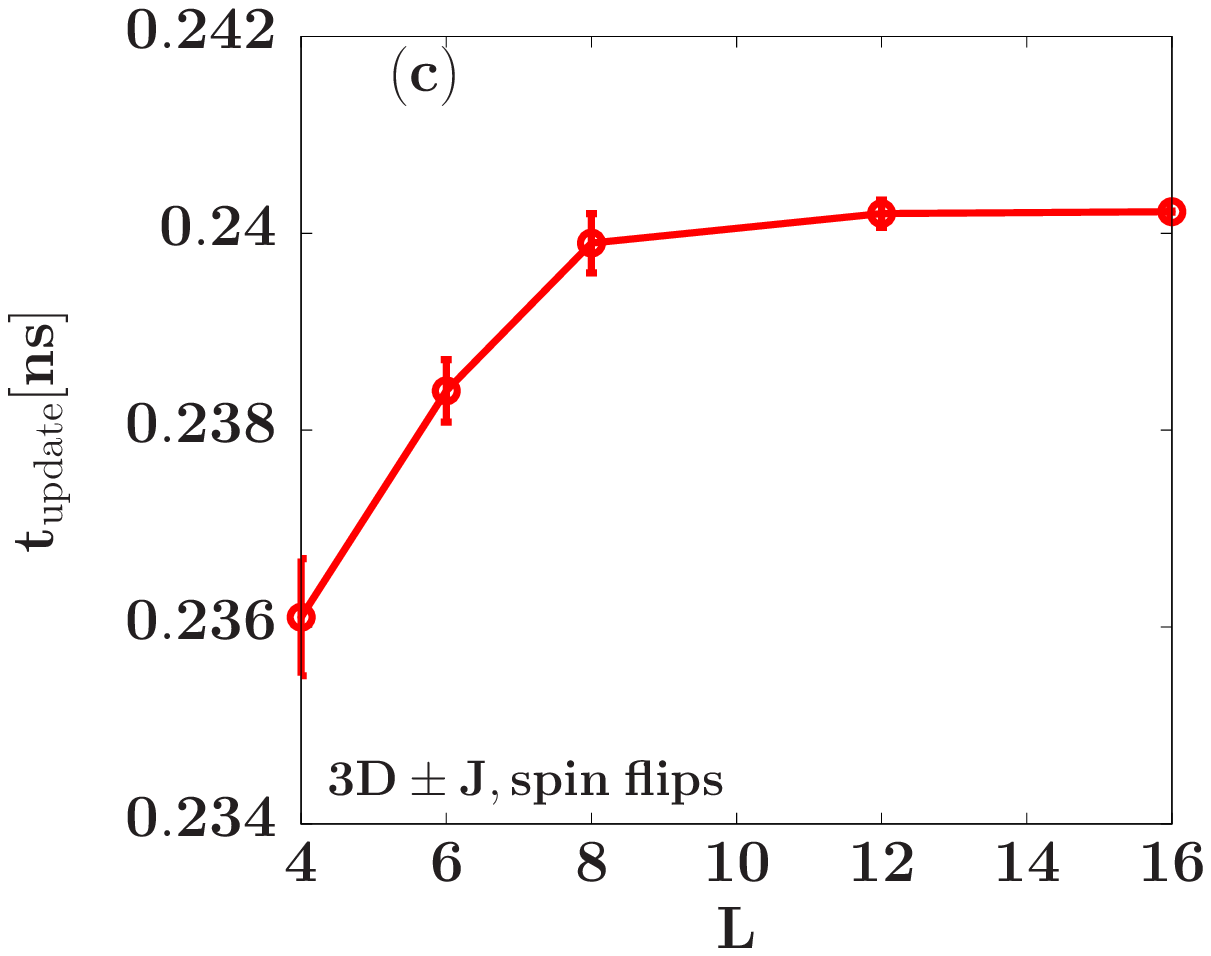}\\
  \includegraphics[scale=0.35]{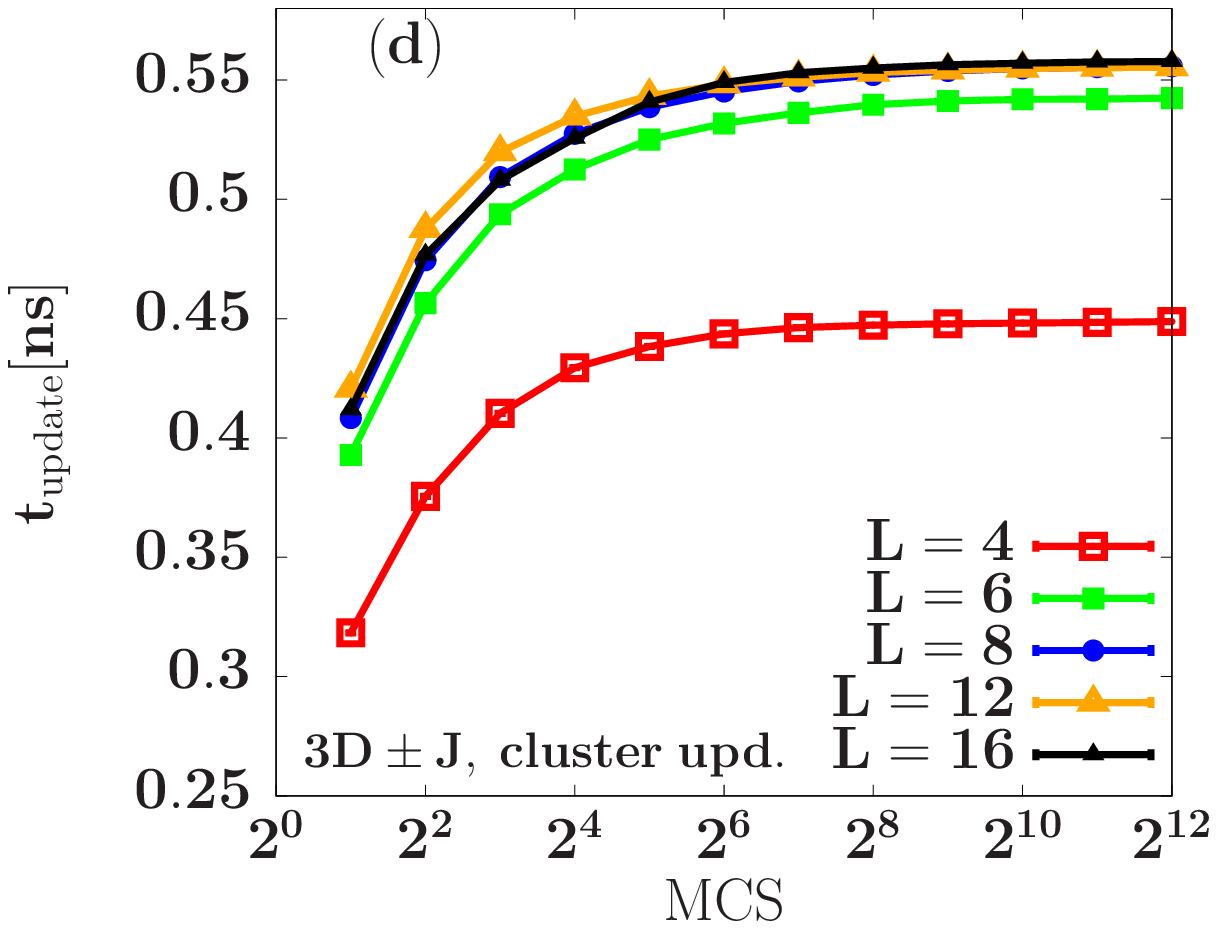}
  \includegraphics[scale=0.35]{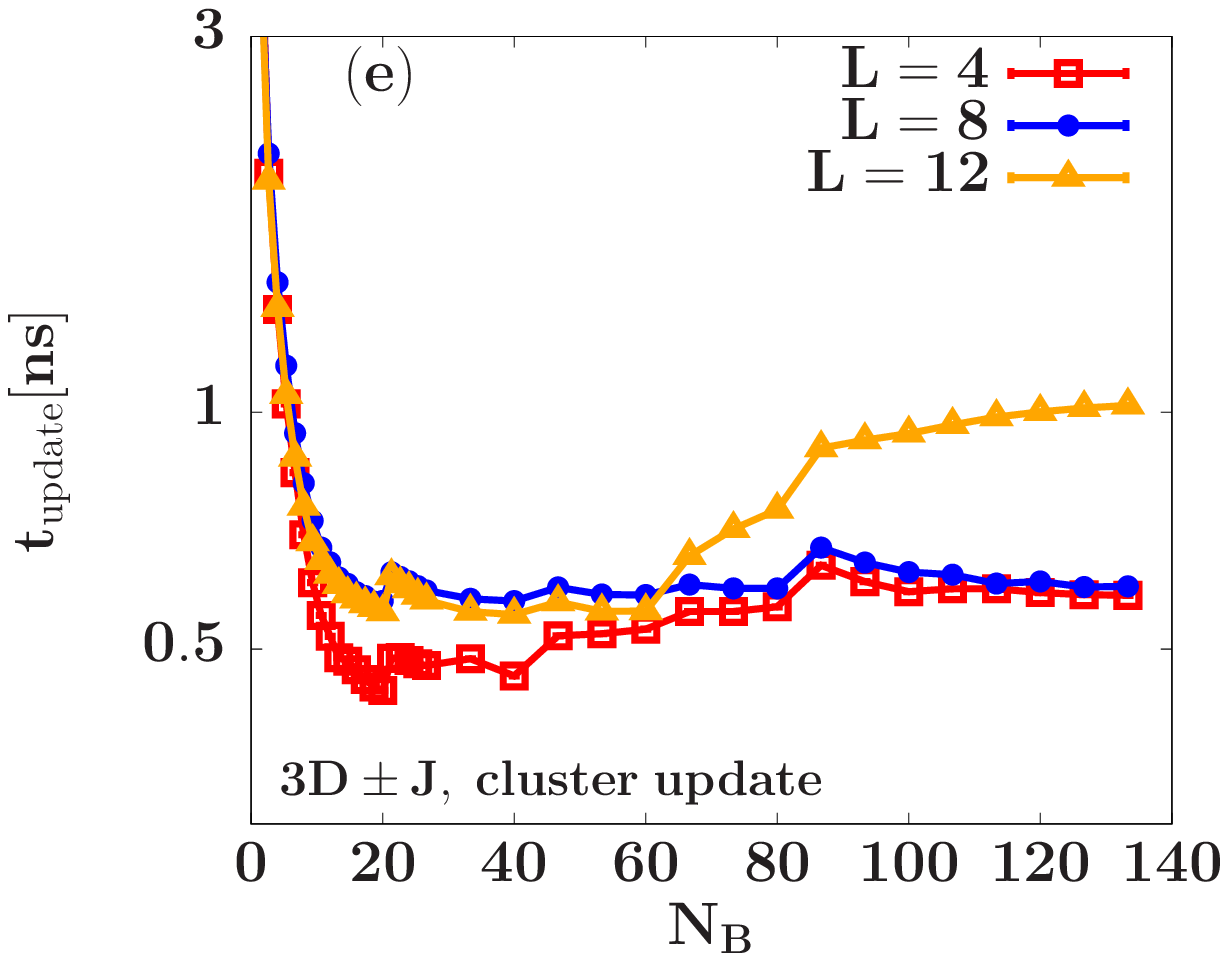}
  \includegraphics[scale=0.35]{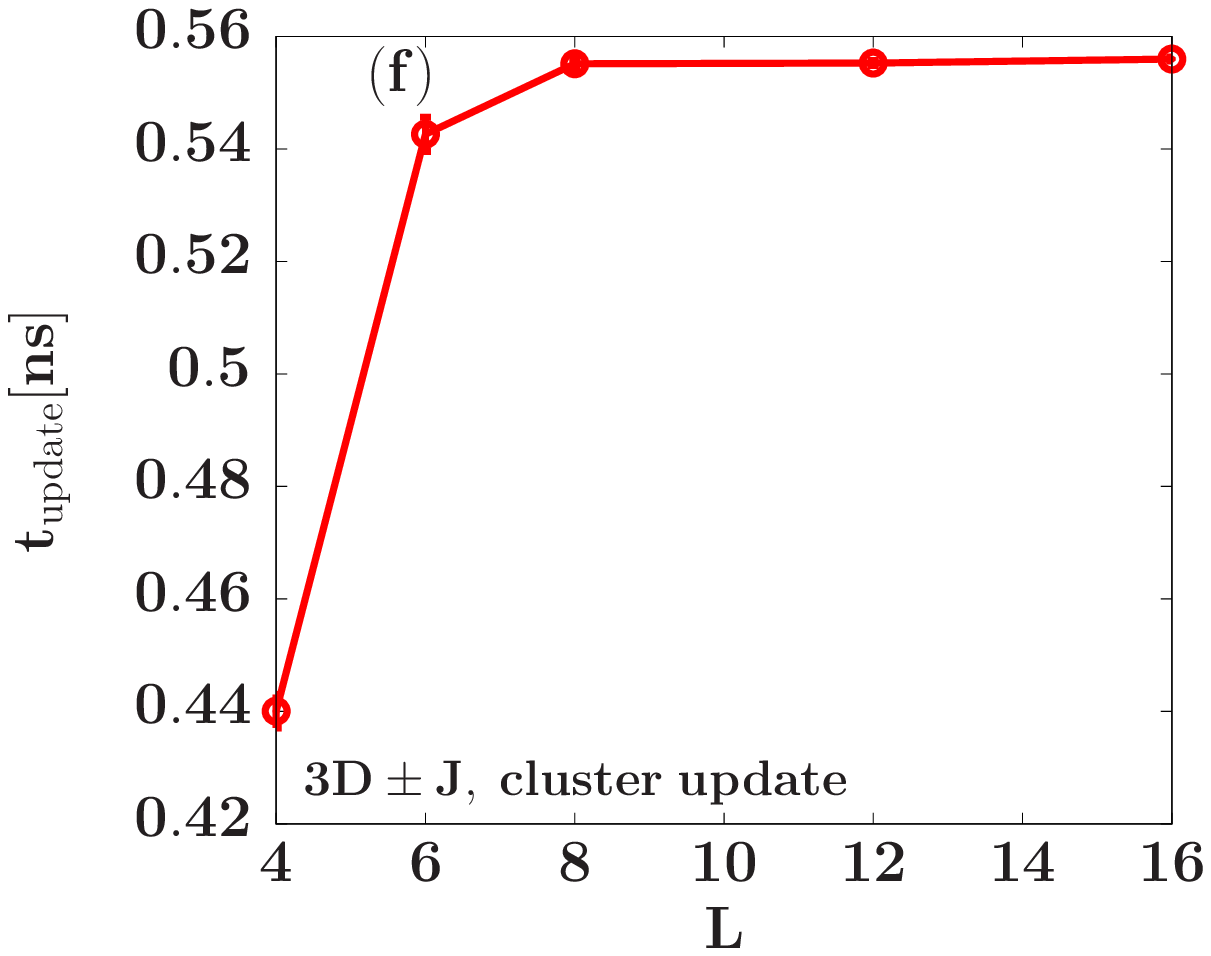}
  \caption{Timing data for GPU simulations of the 3D Edwards-Anderson model with
    bimodal couplings.}
  \label{gpu_perform2}
\end{figure*}

\subsection{Three-dimensional system with bimodal couplings}

We finally also considered the system with bimodal couplings in three dimensions,
where the higher connectivity and related general reduction in memory coalescence
leads to an overall increase in spin-flip times in the Metropolis kernel, cf.\ panels
(a)--(c) of Fig.~\ref{gpu_perform2}. The optimum point of performance at $N_B = 40$
remains unaltered, and the overall single-spin flip times are very nearly independent
of system size, see Fig.~\ref{gpu_perform2} (c). The cluster update, on the other
hand, appears to be performing very differently from the 2D cases, cf.\ panels
(d)--(f). In particular, we observe that the normalized time spent in the
cluster-update kernel does no longer decrease with increasing $L$, but remains
constant beyond $L\gtrsim 8$, see panel (f). This is an effect of the cluster
algorithm itself, however, and not a shortcoming of the presented implementation: at
least without further modifications in the spirit of the proposal put forward in
Ref.~\cite{zhu:15a} the clusters grown by the construction introduced by Houdayer
\cite{houdayer:01} start to percolate at temperatures significantly above the
spin-glass transition point \cite{machta:07,zhu:15a}. As the times shown in
Fig.~\ref{gpu_perform2} (f) are effectively an average over the replicas running at
all the different temperatures considered, the updating times per spin approach a
constant. Here, we do not explicitly attempt to improve the percolation properties of
the update itself, but it is clear that our GPU implementation allows such
modifications without compromising its performance.

\subsection{Cost of the parallel tempering step}
\label{sec:tempering-cost}

\begin{figure}[tb!]
  \centering
  \includegraphics[scale=0.5]{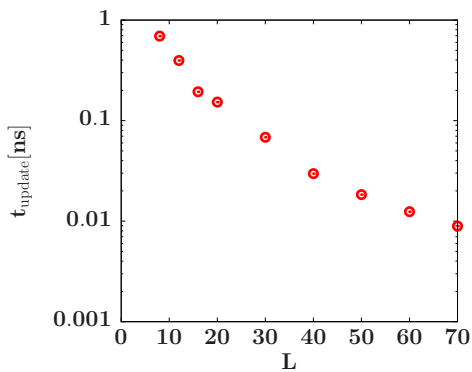}
  \caption {Time spent in the parallel-tempering moves for GPU simulations of the 2D
    bimodal spin-glass model, normalized to the individual spin. The data are
    averaged over 640 disorder realizations.}
  \label{fig:gpu_perform_pt}
\end{figure}

As discussed above in Sec.~\ref{sec:gpu}, the replica-exchange moves in our code are
performed on GPU after transferring the data for the energies from GPU to CPU. While
this might seem wasteful, the simplicity of the resulting code appears very
attractive if only there is no significant performance penalty associated to this
approach. To check whether this is the case we show in Fig.~\ref{fig:gpu_perform_pt}
the updating time for the parallel-tempering step for the 2D bimodal system which,
according to Eq.~\eqref{eq:times} is again normalized to a single spin to make the
result comparable to the data shown in Figs.~\ref{fig:gpu_perform},
\ref{gpu_perform1} and \ref{gpu_perform2}. It is clear that for the larger system
sizes, where most of the resources for a simulation campaign would be invested, the
cost of the replica-exchange steps is small compared to the time spent on updating
spins. While the data shown are specifically recorded for simulations of the 2D
bimodal Edwards-Anderson model, the time taken for the swap moves only depends on the
number of temperature points and it is thus independent of the specific system under
consideration, such that the conclusion of negligible cost of the parallel tempering
holds even more for the computationally more expensive simulations of the 2D Gaussian
and 3D spin-glass models.

\subsection{Speed-up}
\label{sec:comparison}

\begin{figure*}[tb!]
  \centering
  \includegraphics[scale=0.35]{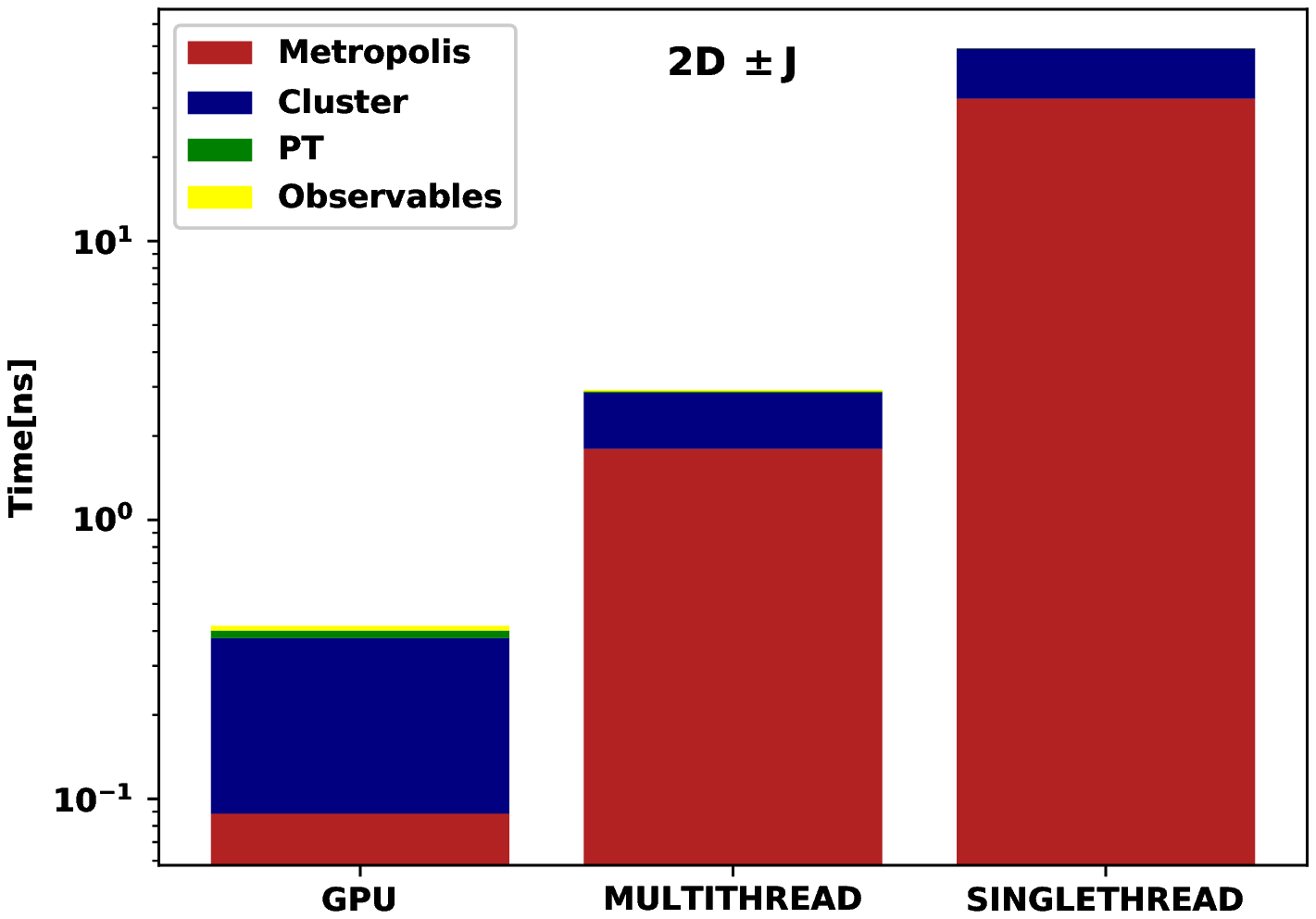}
  \includegraphics[scale=0.35]{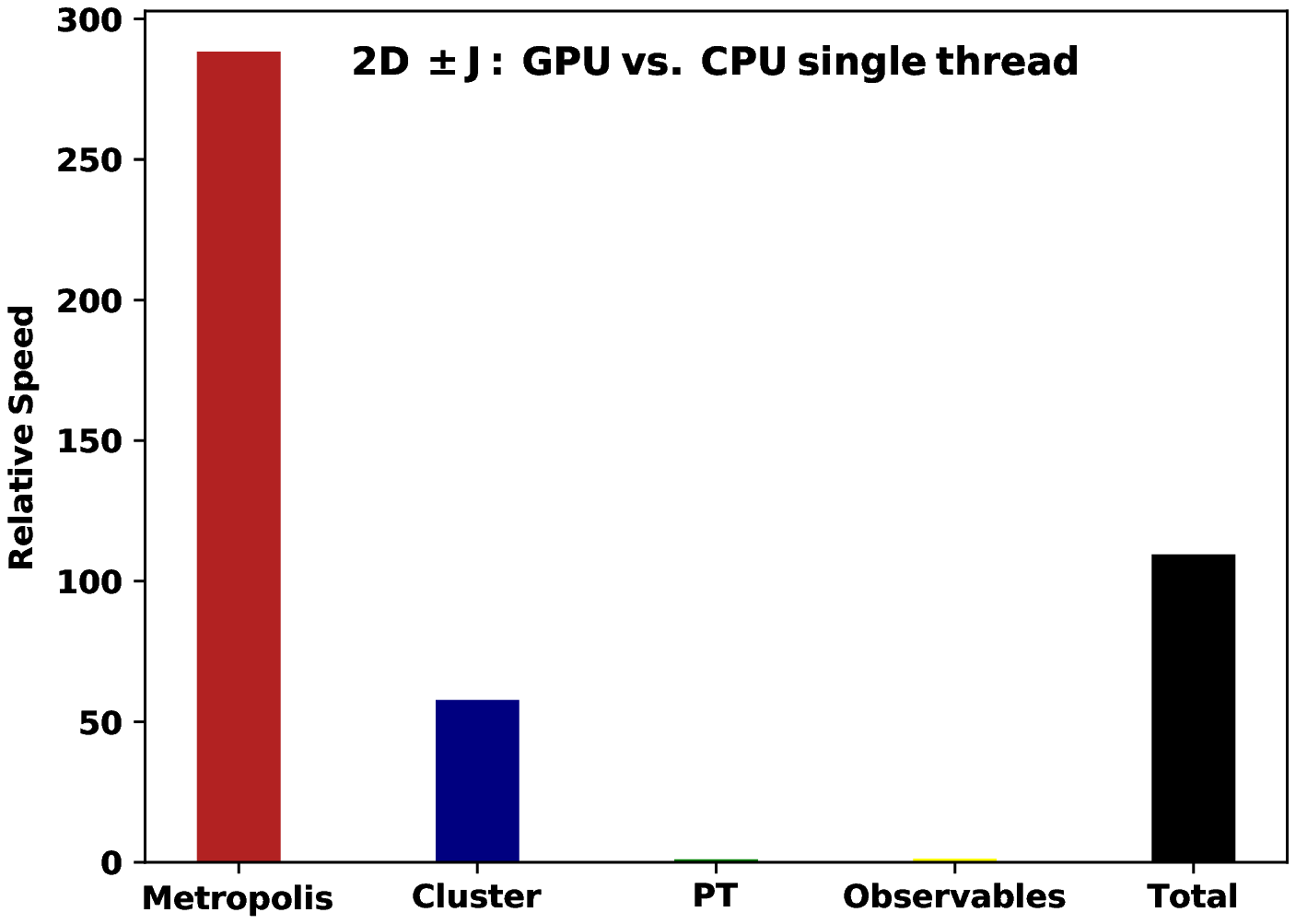}
  \includegraphics[scale=0.35]{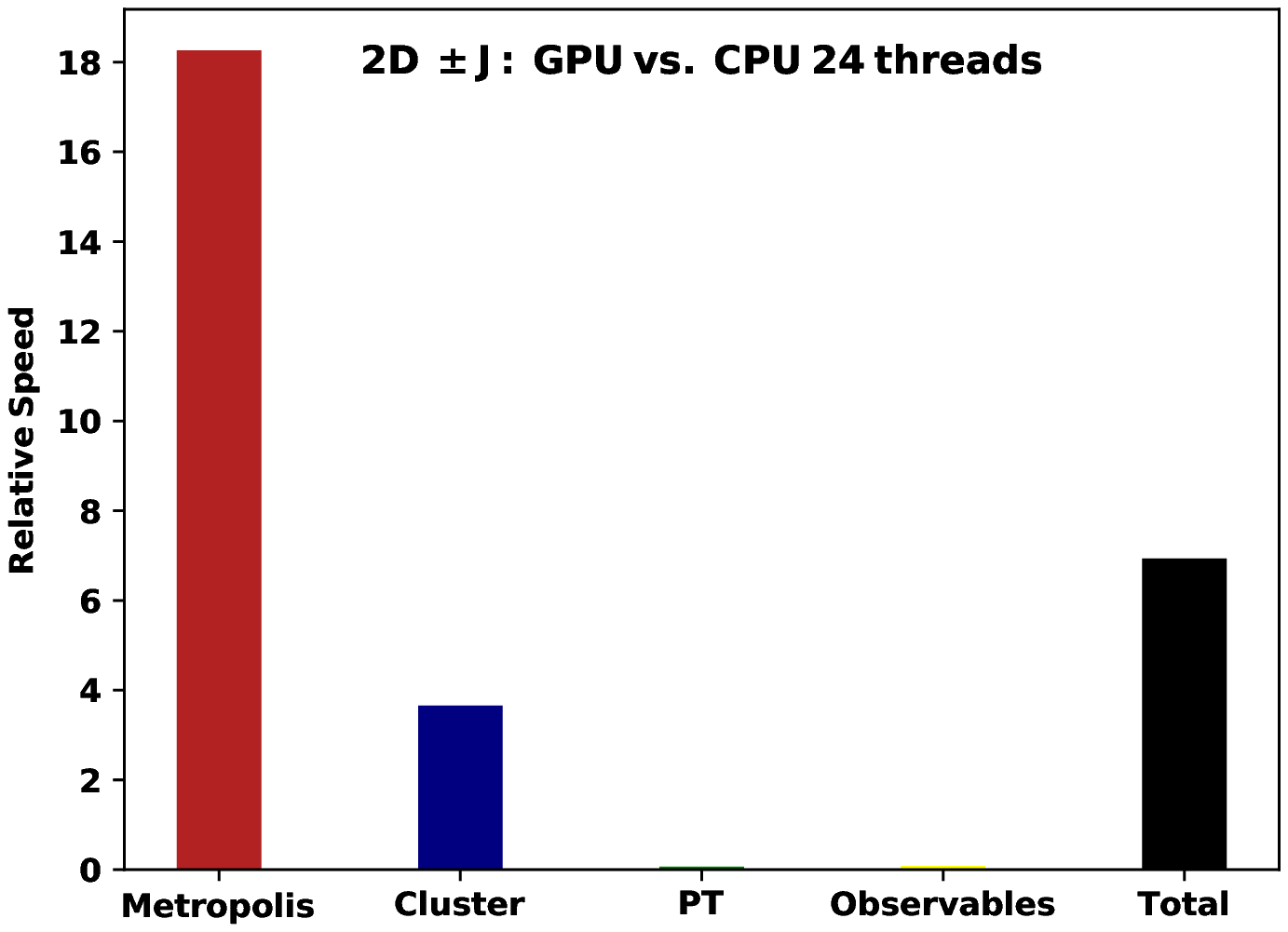}\\
  \includegraphics[scale=0.35]{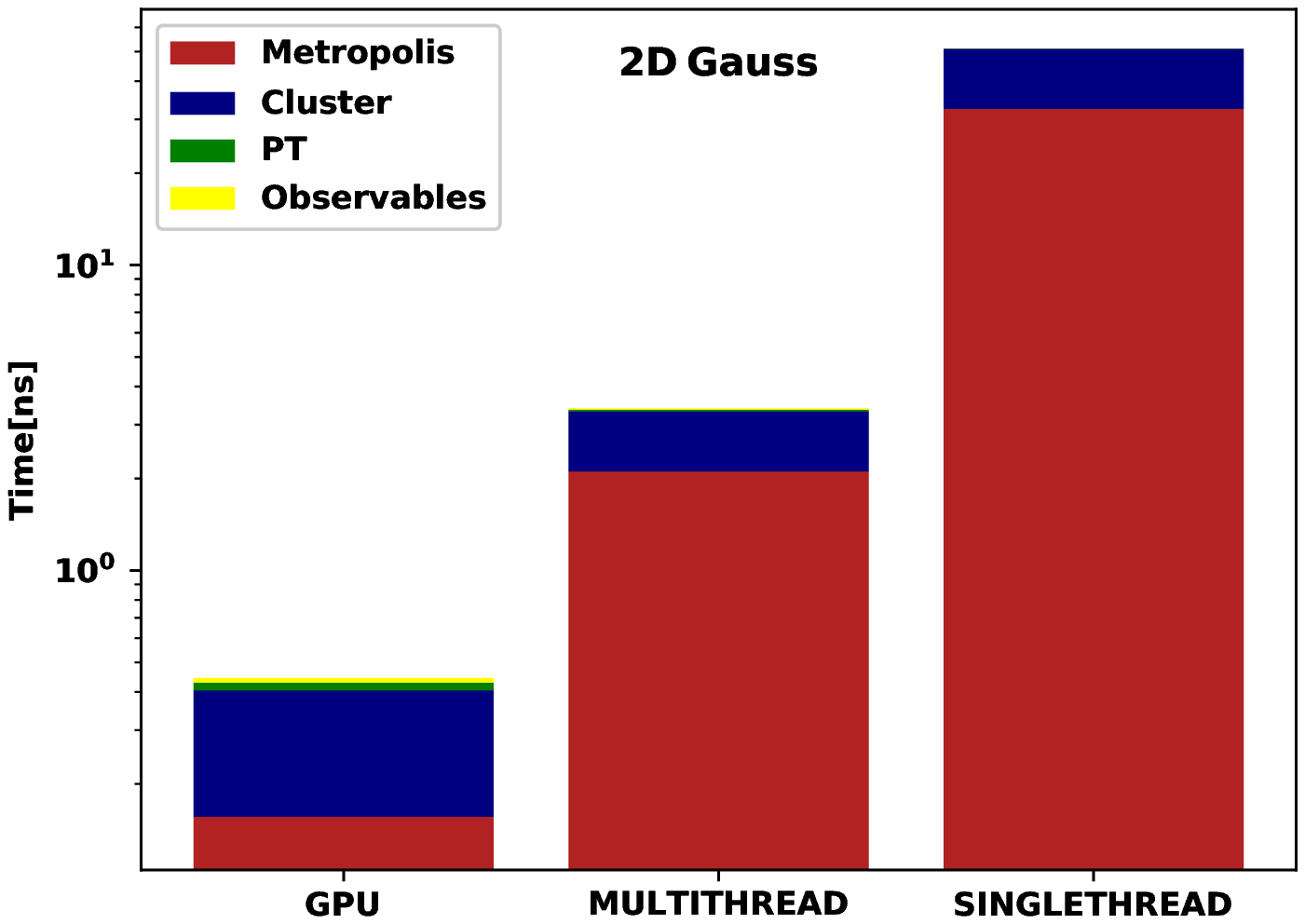}
  \includegraphics[scale=0.35]{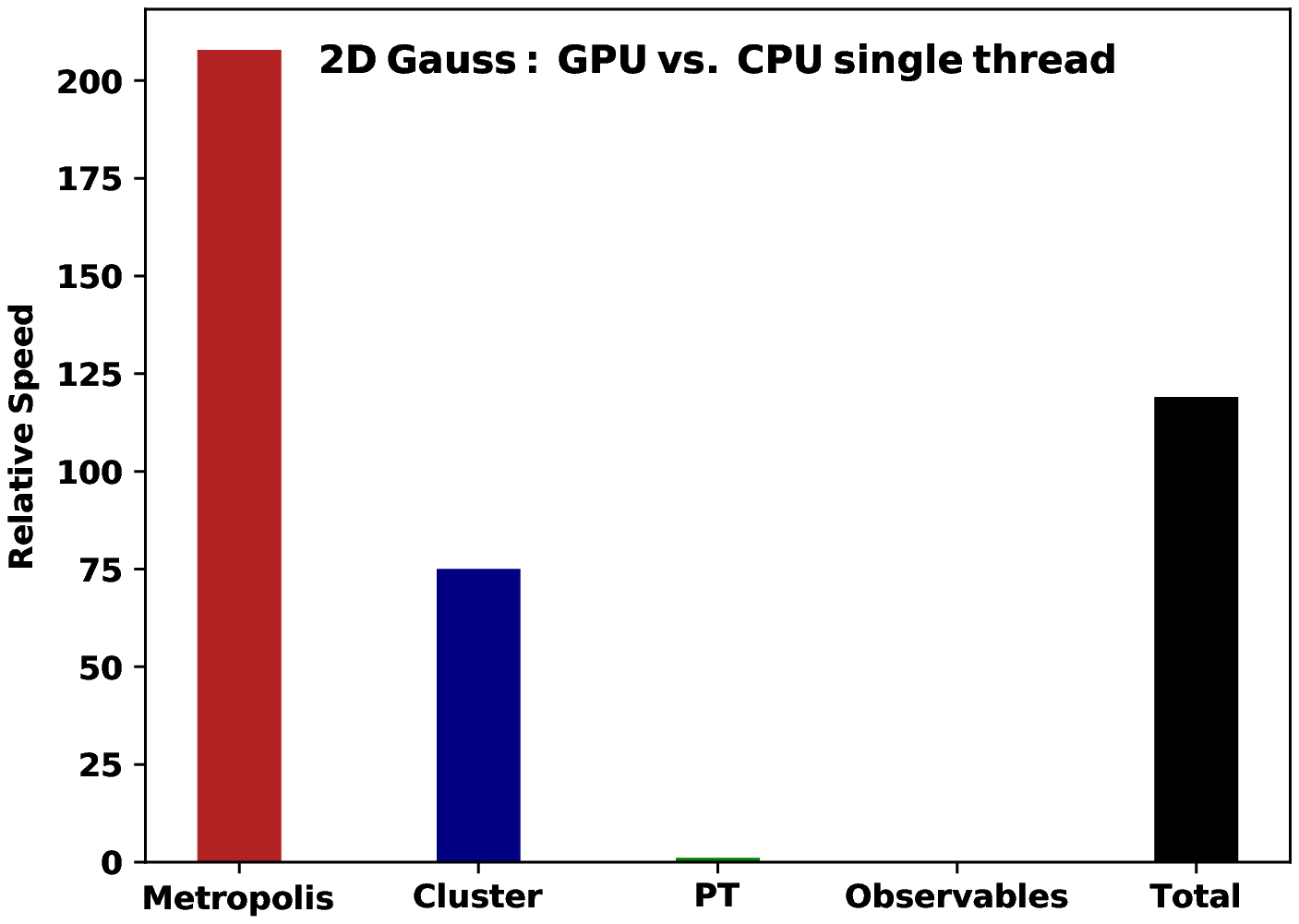}
  \includegraphics[scale=0.35]{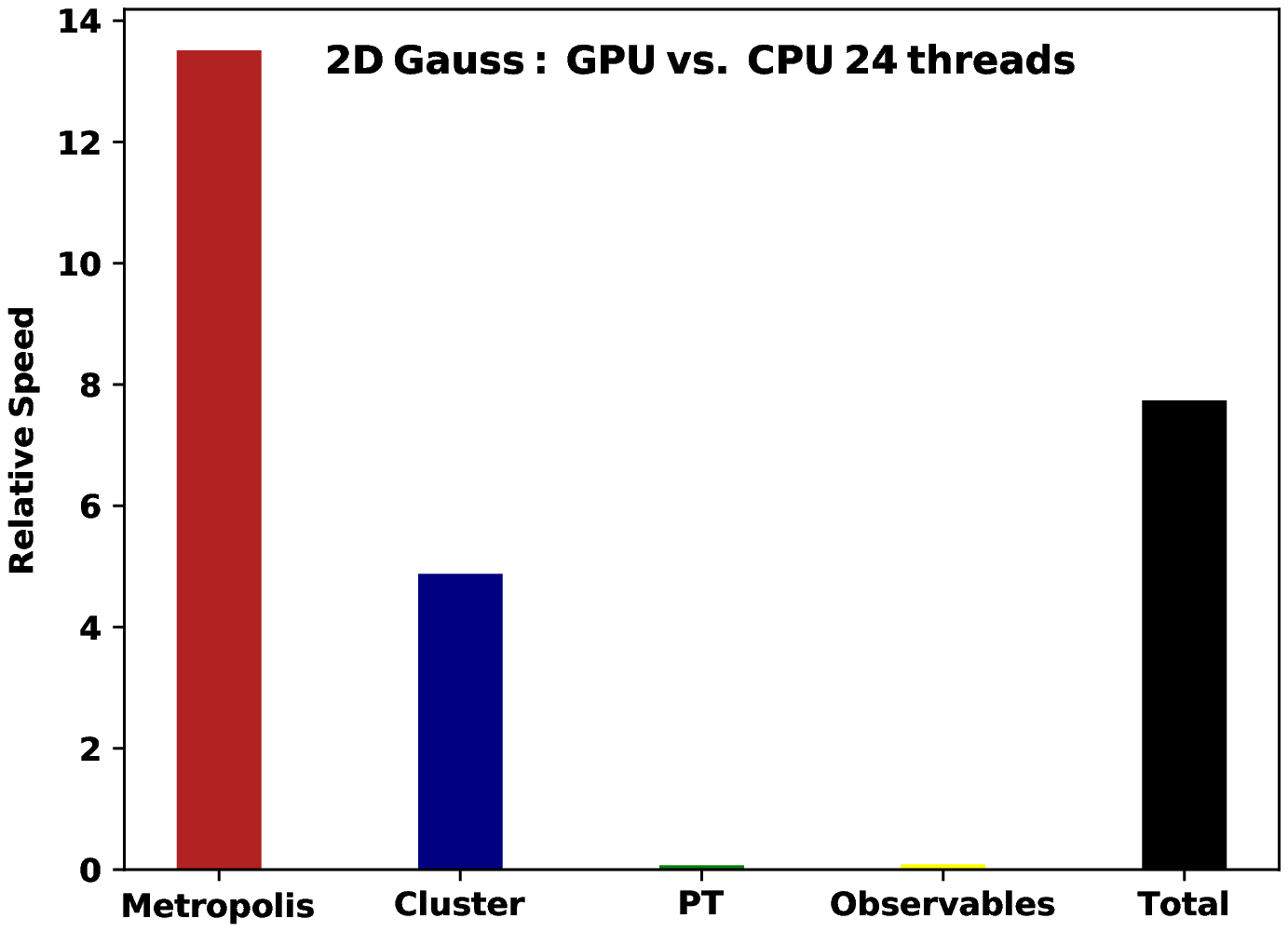}
  \caption{Times spent in different parts of the simulation on a log-scale as compared
    between the GPU and single-threaded as well as multi-threaded CPU implementations
    for 2D systems of linear size $L=40$ drawn from bimodal and Gaussian coupling
    distributions, respectively (left column). The middle and right columns show
    relative timings (speed-ups) of the GPU code as compared to the single-threaded
    (middle column) and multi-threaded (right column) CPU runs, respectively.}
  \label{speed-up}
\end{figure*}

We finally turn to a comparison of the performance of the GPU code introduced above
to a reference CPU implementation of exactly the same simulation. The results of this
comparison are summarized in Fig.~\ref{speed-up}. To better understand the
performance of different parts of the simulation, we broke the total run-time of both
the GPU and CPU simulations into the times spent in (1) flipping spins using the
Metropolis algorithm, (2) performing Houdayer's cluster update, (3) exchanging
replicas in the parallel tempering method, and (4) in measurements of the basic
observables. The resulting break-down of times is illustrated in the left column of
Fig.~\ref{speed-up}, showing the two-dimensional bimodal (top) and Gaussian (bottom)
models, respectively. We find the system size dependence of the times per spin to be
rather moderate beyond a certain system size (cf.\ the representations in
Figs.~\ref{fig:gpu_perform} and \ref{gpu_perform1}), and consequently we only show
results for a single size $L=40$ in Fig.~\ref{speed-up}. The CPU code was run on a
6-core/12-threads Intel Xeon E5-2620 v3 CPU running at 2.40GHz either using a single
thread or running 24 disorder realizations in parallel in the multi-threaded version
on a dual-CPU system using hyper-threading, while the GPU code as before was
benchmarked on the GTX 1080. It is clearly seen that the CPU code spends the majority
of time in flipping spins, whereas on GPU more time is spent in the cluster update as
this does not parallelize as well as the Metropolis algorithm. To support this
observation further, we show the relative run-times, i.e., speed-ups, of the
individual parts of the simulation as compared to the single-threaded (middle column)
and multi-threaded (right column) CPU runs, respectively. While we observe speed-ups
between GPU and a single CPU thread of between 200 and 300 for the spin flips, the
cluster update only improves by a factor of 50--75, on average. The parallel
tempering moves, on the other hand, are executed on CPU even in the GPU version of
the code and do not experience any speed-up, but their contribution to the overall
run-time is negligibly small. Overall, we still observe a speed-up by a factor of
about 125 for the whole GPU simulation compared to a single-threaded CPU code. For
the multi-threaded CPU implementation run-times are divided by a factor in between
the number 12 of physical cores and the number 24 of threads, which indicates good
intra-CPU scaling. As a result, compared to a full dual-CPU node our GPU
implementation achieves an about eight-fold speed-up overall.

\section{Summary}
\label{sec:conclusion}

We have discussed the computational challenges of simulating disordered systems on
modern hardware, and presented a versatile and efficient implementation of the full
spin-glass simulation stack consisting of single-spin flips, cluster updates and
parallel-tempering updates in CUDA.  Due to the favorable relation of performance to
price and power consumption in GPUs, they have turned into a natural computational
platform for the simulation of disordered systems. While a range of very efficient,
but also very complex, simulational codes for the problem have been proposed before
\cite{bernaschi:10,yavorskii:12,baity-jesi:14,lulli:15}, our focus in the present
work was on the provision of a basic simulation framework that nevertheless achieves
a significant fraction of the peak performance of GPU devices for the simulation of
spin-glass systems.  To be representative of typical installations accessible to
users, we used Nvidia GPUs from the consumer series (GTX 1080).

Comparing our GPU implementation to our reference CPU code, we find a speed-up factor
of more than 200 for the Metropolis kernel, which is not far from the performance of
more advanced recent GPU simulations of spin models, see, e.g.,
Ref.~\cite{barash:16}. The cluster update algorithm used here is much less well
suited for a parallel implementation, especially since it is a single-cluster variant
for which the strong variation in cluster sizes leads to the presence of idling
threads. Here, performance could be further improved by moving to a multi-cluster
code, in line with the experience for ferromagnets \cite{weigel:10b}. Nevertheless,
overall we still observe a speed-up of the GPU code of about 125 as compared to a
single CPU thread, and of about 8 as compared to a full dual-processor CPU node with
24 threads.

In combination with our simple, parametric scheme for choosing the temperature
schedule, the proposed simulation framework provides an accessible and highly
performant code base for the simulation of spin-glass systems that can easily be
extended to other systems with quenched disorder such as the random-field problem
\cite{kumar:18}.

\begin{acknowledgements}
  We would like to thank Jeffrey Kelling for fruitful discussions. Part of this work
  was financially supported by the Deutsche Forschungsgemeinschaft (DFG, German
  Research Foundation) under project No.\ 189\,853\,844 -- SFB/TRR 102 (project B04)
  and under Grant No. JA~483/31-1, the Leipzig Graduate School of Natural Sciences
  ``BuildMoNa'', the Deutsch-Franz\"osische Hochschule (DFH-UFA) through the Doctoral
  College ``${\mathbb L}^4$'' under Grant No. CDFA-02-07, the EU through the IRSES
  network DIONICOS under contract No. PIRSES-GA-2013-612707, and the Royal Society
  through Grant No.\ RG140201.
\end{acknowledgements}


\end{document}